\documentclass[aps,prx,groupedaddress,twocolumn,showpacs,longbibliography]{revtex4-1}
\usepackage{amsmath}
\usepackage{amsthm, amssymb}    
\usepackage{braket}              
\usepackage{graphicx,tikz}
\usepackage{color}
\usepackage{times, verbatim}      
\usepackage{mathtools}
\usepackage{natbib}
\usepackage{hyperref}
\hypersetup{
       colorlinks=true,
}
\usepackage{etoolbox}
\newtoggle{includeHubbardDetails}
\togglefalse{includeHubbardDetails}

\def \tr{{\mbox{tr~}}}

\def \be{\begin{equation}}
\def \ee{\end{equation}}
\def \bea{\begin{eqnarray}}
\def \eea{\end{eqnarray}}

\newcommand{\cpn}{\mathbb{C}P^n}
\newcommand{\im}{\operatorname{Im}}
\newcommand{\re}{\operatorname{Re}}
\begin{document}
\title{Beyond the Berry Phase: Extrinsic Geometry of Quantum States}

\author{Alexander Avdoshkin$^{1}$}
\email{alexander\_avdoshkin@berkeley.edu}
\author{Fedor K. Popov$^{2}$}
\email{fpopov@nyu.edu}
\affiliation{ {
1. Department of Physics, University of California, Berkeley, CA 94720, USA \\
2. CCPP, Department of Physics, NYU, New York, NY, 10003, USA
  }}

\begin{abstract}
{
Consider a set of quantum states $| \psi(x) \rangle$ parameterized by $x$ taken from some parameter space $M$. We demonstrate how all geometric properties of this manifold of states are fully described by a scalar gauge-invariant Bargmann invariant $P^{(3)}(x_1, x_2, x_3)=\operatorname{tr}[P(x_1) P(x_2)P(x_3)]$, where $P(x) = |\psi(x)\rangle \langle\psi(x)|$. Mathematically, $P(x)$ defines a map from $M$ to the complex projective space $\cpn$ and this map is uniquely determined by $P^{(3)}(x_1,x_2,x_3)$ up to a symmetry transformation. The phase $\arg P^{(3)}(x_1,x_2,x_3)$ can be used to compute the Berry phase for any closed loop in $M$, however, as we prove, it contains other information that cannot be determined from any Berry phase. When the arguments $x_i$ of $P^{(3)}(x_1,x_2,x_3)$ are taken close to each other, to the leading order, it reduces to the familiar Berry curvature $\omega$ and quantum metric $g$. We show that higher orders in this expansion are functionally independent of $\omega$ and $g$ and are related to the extrinsic properties of the map of $M$ into $\cpn$ giving rise to new local gauge-invariant objects, such as the fully symmetric 3-tensor $T$. Finally, we show how our results have immediate applications to the modern theory of polarization, calculation of electrical response to a modulated field and physics of flat bands.

 }
\end{abstract}
\maketitle
%

\section{Introduction}

Since its introduction in \cite{Berry1984} the Berry phase has become an indispensable tool in the study of quantum systems and has even been extended to classical system as the Hannay angle \cite{Hannay_1985}. Initially introduced as the geometric contribution that to the phase accumulated under an adiabatic evolution, the Berry phase is directly manifested in some phenomena, e.g. the Aharonov-Bohm effect. Although its applications has proved most fruitful in condensed matter physics\cite{Resta_2000, RevModPhys.82.1959} where it has been instrumental in understanding the role of topology, in particular, in explaining the quantization of the quantum Hall effect \cite{shapere1989geometric}. The related object of Berry connection has also been directly related to the polarization of crystalline systems.


Another geometric characteristic of a set of states, the quantum metric, was introduced in \cite{Provost1980} as the infinitesimal version of the quantum distance $D(\psi, \chi)$ that, for any two states $|\psi\rangle$ and $|\chi\rangle$, is defined as $ D(\psi, \chi) = 1 - |\langle \psi| \chi \rangle|^2$. The quantum metric has found uses in detecting phase transitions\cite{Wang_2015, Campos_Venuti_2007, PhysRevLett.99.100603, GU_2010}, non-adiabatic responses \cite{KOLODRUBETZ20171} and quantum chaos\cite{PhysRevX.10.041017}. 

More recently the Berry curvature and quantum metric have arose as two complementary objects that capture the geometry of an electronic band structure in solids. This supplements the energy spectrum that has been traditionally used as the main characteristic of a material. The most notable examples of the uses of geometry are non-linear optics \cite{Ahn_2020, ahn2022riemannian}, nonreciprocal directional
dichroism \cite{Gao_2019}, fraction quantum Hall insulators \cite{parameswaran2013fractional}, flat band superconductivity \cite{peotta2015superfluidity}, anomalous Hall conductivity\cite{Kozii_2021} and the theory of the insulating state \cite{resta2011insulating}.

Central to any discussion of the geometry of a manifold of states is the idea of gauge invariance --- states $|\psi\rangle$ and $e^{i \phi}|\psi\rangle$ are physically equivalent. Consequently, if we have a manifold of states $|\psi(x)\rangle$, where $x$ is a parameter taken from some manifold $M$, the geometry of $|\psi(x)\rangle$ is equivalent to the geometry of  $e^{i\phi(x)}\ket{\psi(x)}$. This change of state representatives is referred to as a gauge transformation. Thus, any quantity that purports to characterize the geometry of $|\psi(x)\rangle$ must remain unchanged under all gauge transformations. Naturally, the Berry curvature and quantum metric 
are gauge invariant.

In this work, we propose an explicitly gauge invariant way to fully describe the geometry of any set of states $|\psi (x)\rangle$.
The mathematical way to incorporate gauge invariance is to represent quantum states as elements of the complex projective space $\cpn$ (see Appendix. \ref{sec:cpn}) instead of the Hilbert space $\mathcal{H}$.
Thus, formally, the object of our study are maps from $M$ to $\cpn$. Practically, the mapping to $\cpn$ is realized by working with the projectors $P(x)=\ket{\psi(x)}\bra{\psi(x)}$. Any object constructed out of a combination of these projectors is automatically gauge-invariant. We will show that all geometric information about the map $M \to \cpn$ is encoded in the Bargmann invariant \cite{bargmann1964note}  $P^{(3)}(x,y,z) = \tr\left[P(x)P(y)P(z)\right]$, that we will also refer to as the three-point function. The Bargmann invariant was originally introduced to aid in the proof of Wigner's theorem on symmetry operators, but later found uses in quantum information science \cite{bengtsson2017geometry} and quantum optics \cite{simon1993bargmann}. 
 We will show how to express all known geometric structures via three-point functions and also explain that, generally, $P^{(3)}$ contains information that cannot be recovered from even the global knowledge of the Berry phase and quantum metric.


More precisely, the Bargmann invariant contains novel information 
when the parameter manifold has dimension less than the dimension of $\cpn$. To make this clear, in our work we use the interpretation of the 3-point functions $P^{(3)}(x,y,z)$ in terms of the geodesic triangle build on the images of $x,y$ and $z$ in $\cpn$\cite{hangan1994geometrical,mukunda1993quantum, pancharatnam1956generalized}. When $\dim M < \dim \cpn$, the image of $M$ forms a proper submanifold within $\cpn$ and the geodesics between different points of $M$ generically will not be contained within $M$, thus. revealing extrinsic geometry. We identify a fully symmetric 3-tensor $T_{\alpha \beta \gamma}$ as the simplest local object that captures this novel geometric information.

Previously, the approach of describing a quantum state depending on a parameter as a map to $\cpn$ has been successfully applied to classifying topological insulators \cite{Chiu_2016}. In that case only the homotopy class of the map $M \to \cpn$ mattered and the approach of the present work can be seen as an extension of the topological approach to also include geometric properties of the map. 

Beside capturing new geometry beyond the Berry curvature and quantum metric, our approach also significantly simplifies certain computations that involve band structures, yielding explicitly gauge invariant answers and revealing the geometric objects behind physical quantities. We will illustrate that by computing the conductivity of a flat band material and the polarization distribution in an insulator. Additionally, we explain what our results imply when applied to the geometry of Landau levels and the band structure of twisted bilayer graphene in the chiral limit.

The remainder of the paper is organized as follows. In Section \ref{sec:projectors} we introduce the 3-point function $P(x,y,z)$ and describe its basic properties. In the next Section \ref{sec:berry_phase}, the Berry phase is reinterpreted via $P(x,y,z)$ and its infinitesimal version $\mathcal{A}^{x}_{\alpha}(y)$. Section \ref{sec:int_vs_ext} describes the distinction between intrinsic and extrinsic geometry that plays an important role in this work. Section \ref{sec:local} covers the local geometric invariants that arise in the expansion of $P(x,y,z)$ when its arguments are close to each other. Section \ref{sec:holomorphic} describes the special properties of holomorphic maps $M \to \cpn$ and Section \ref{sec:bands} shows how the approach of this work applies to band structure calculations. Finally, in Section \ref{sec:disuccsion} we conclude with a discussion and an outlook. 

\section{Geometry via projectors} \label{sec:projectors} 
Assume we have a set of wave functions, $|\psi(x)\rangle$, where $x$ belongs to some parameter space $M$, normalized by $\langle \psi(x)|\psi(x)\rangle = 1$. We notice that the projector on the given state $P(x) = |\psi(x)\rangle \langle\psi(x)|$ is invariant under $|\psi(x)\rangle \to e^{i \phi(x)} |\psi(x)\rangle$ and, consequently, all expressions involving projectors will be gauge invariant. Likewise, any hermitian operator $\mathcal{O}$ can be represented via projectors: if the eigenvalues of $\mathcal{O}$ are $\lambda_i$ and the eigenvectors are $| i\rangle$ we have
\bea
\mathcal{O} = \sum_i \lambda_i P^{\mathcal{O}}_i,
\eea
where $P^{\mathcal{O}}_i = |  i\rangle\langle i | $.
With that, any physical observable can be represented as a trace of a product of projectors. For example, for a single operator $\mathcal{O}$:
\bea
\langle \psi(x) | \mathcal{O}| \psi(x) \rangle = \sum_i \lambda_i \tr[P(x) P^{\mathcal{O}}_i] 
\eea
and, similarly, for a pair of operators $\mathcal{O}_1, \mathcal{O}_2$:
\bea \label{eq:tro}
\langle \psi(x) | \mathcal{O}_1 \mathcal{O}_2| \psi(x) \rangle = \sum_{i,} \lambda_i \lambda_j \tr[P(x) P^{\mathcal{O}_1}_i P^{\mathcal{O}_2}_j].
\eea
With that one can convince oneself, that the basic building block for constructing any physical observable is the following $k$-point function \footnote{Additionally, one might need to enlarge the parameter space $M$ to include the eigenstates of operators in Eq. \eqref{eq:tro}}
\be \label{eq:traces}
P^{(k)}(x_1, x_2, \dots, x_k) = \tr [P(x_1) P(x_2) \cdots P(x_k)],
\ee
where the superscript $(k)$ will be omitted later in the text if it does not lead to ambiguities.  Eq. \eqref{eq:traces} is invariant under a circular permutation of its arguments: $P(x_1, x_2, \dots, x_k) = P(x_k, x_1 \dots, x_{k-1})$.  Because of the property $P^2 = P$ 
if two consecutive arguments of an $n$-point function coincide it reduces to an $n-1$-point function: $P^{(n)}(x_1, \dots, x, x, \dots, x_k) = P^{(n-1)}(x_1, \dots, x, \dots, x_k) $. Furthermore, any $n$-point function can be expressed via two-and three-point functions $P(x,y)$ and $P(x,y,z)$:
\begin{gather}
    \tr \left[P_1 P_2 \ldots P_{k-1} P_k P_{k+1} \ldots P_n\right] = \notag\\
    \tr \left[P_1 P_2 \ldots P_{k-1} P_{k+1} \ldots  P_n\right] \frac{\tr\left[P_{k-1} P_k P_{k+1}\right]}{\tr \left[P_{k-1} P_{k+1}\right]} \label{eq:red},
\end{gather}
therefore it is sufficient to only study the Bargmann invariant $P(x,y,z)$ to get the information about all $n$-point functions.

    




Next, we argue that the knowledge of $P(x_1,x_2,x_3)$ for all values of $x_i$ allows one to uniquely determine all gauge-invariant properties of $| \psi (x) \rangle$ and, in principle, compute all physical observables. This fact is a corollary of the the results in Appendix \ref{app:gps_theorem}, where it is shown that any $n+1$ points of general position in $\cpn$ can be reconstructed from their relative 3-point functions up to a holomorphic rigid motion (change of basis in the original Hilbert space) and these $n+1$ points can be used a reference set to locate all other points (the GPS theorem). 

\begin{figure}
    \begin{tikzpicture}[scale=1.5]
    \node at (-0.5,0.25) {$M$};
    \node at (2.3,1.5) {$\mathbb{C}P^n$};
    \node at (0.6,1.7) [above] {$\gamma$};
    \node at (0.35,0.25) {$\tilde{\gamma}$};
       \draw (0.1,0.5)..controls (0.35,0.3) and (0.85,0.7)..(1.1,0.5);
       \draw (-1,0)--(0,1)--(2,1)--(1,0)--cycle;
       \draw[thick] (0.1,0.5)..controls (0.6,2)..(1.1,0.5);
       
    \begin{scope}[xshift = 75]
     \draw (-1,0)--(0,1)--(2,1)--(1,0)--cycle;
     \draw (-0.2,0.3)..controls (0.35,0.1)..(1.1,0.3) .. controls (1.0,0.7) .. (0.55,0.8) .. controls (0,0.65) .. cycle;
     \draw[thick] (-0.2,0.3)..controls (0.45,0.7)..(1.1,0.3);
     \draw[thick] (1.1,0.3) .. controls (1.1,1.6) .. (0.55,0.8); 
     \draw[thick] (0.55,0.8) .. controls (0,1.65) .. (-0.2,0.3);
     \node at (-0.5,0.25) {$M$};
  \end{scope}
    \end{tikzpicture}
  \caption{On the left: The difference between sumanifold and global geodesics is illustrated. The path $\tilde{\gamma}$ is the geodesic subject to the submanifold M and $\gamma$ is the global geodesic in the ambient manifold. On the right: The geodesic triangle in the ambient $\cpn$ that captures the information about the three-point function $P(x,y,z)$ is drawn with bold, its intrinsic counterpart within $M$ is also shown.}
   \label{fig:geodesicsovermanifold}
\end{figure}
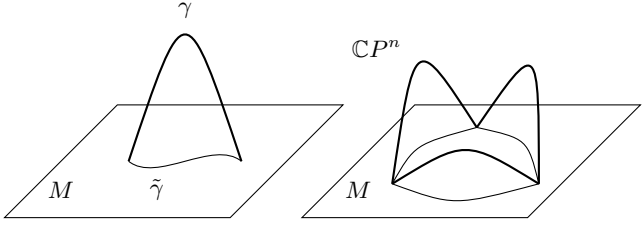

After having established the fundamental role played by $P(x,y,z)$ we turn to describing its properties. The most familiar object arises when we take two of the arguments to be the same and obtain a two point function:
\begin{gather*}
P(x,x,y) = P(x,y) = |\langle \psi(x) | \psi(y) \rangle|^2=\\
= 1 - D^2(x,y) = \cos^2 d(x,y) 
\end{gather*}
here $D(x,y)$ is the quantum distance between states $| \psi(x) \rangle$, $| \psi(y) \rangle$ and $d(x,y)$ is the length of the shortest geodesic that connects the points corresponding to $| \psi(x) \rangle$ and $| \psi(y) \rangle$ in $\cpn$, see Appendix \ref{app:3pt}. This geodesic does not necessarily belong to the submanifold of states $|\psi(x)\rangle$, $x\in M$, see Fig.\ref{fig:geodesicsovermanifold}, which is the reason why extrinsic properties of the map $M \to \cpn$ appear in certain observables. However, since for small distances the geodesic stays close to the submanifold, the first few orders in the expansion are captured by just the intrinsic geometry.

Let us now turn to the general position of $x,y$ and $z$ in $P(x,y,z)$. The absolute value of the three point function is still related to the quantum distance:
\begin{gather}
|P(x,y,z)| = \sqrt{P(x,y) P(y,z) P(z,x)}.
\end{gather}
At the same time, the phase of $P(x_1,x_2,x_3)$ is given by the flux of the $\cpn$ symplectic form (see Appendix \ref{sec:cpn} for the definition) through the geodesic triangle $\gamma_{123}=\gamma_{12}+\gamma_{23}+\gamma_{31}$ (see Appendix \ref{app:3pt} for the  proof) build on $x_1$, $x_2$ and $x_3$ within $\cpn$ (here $\gamma_{ij}$ is an oriented geodesic connecting the points $x_i$ and $x_j$):
\bea
\Phi(x,y,z) = -\arg P(x,y,z) = \int\limits_{\gamma_{123}} A,
\eea
which we visualize in Fig.\ref{fig:geodesicsovermanifold}. 
Analogously to the absolute value, the phase is also sensitive to the extrinsic geometry when the geodesic triangle is not contained within the submanifold. In the next section, we will describe in detail how the phase of the three point function is related to the Berry phase.

\section{Berry phase as a gauge-invariant integral} \label{sec:berry_phase} 
For a closed loop in the parameter space $M$ the Berry phase is defined as
\bea \label{eq:berry_def}
\phi(\gamma) = \int_{\gamma} A,
\eea
where
\bea \label{eq:a_mu}
A_{\mu}(x) = i \langle \psi(x) | \partial_{\mu} \psi(x) \rangle
\eea
is the Berry connection. $A_{\mu}(x)$ is not gauge invariant and transforms into $A_{\mu}(x) + \partial_{\mu} \phi(x)$ under $|\psi(x)\rangle \to e^{i\phi(x)}|\psi(x)\rangle$, but Eq. \eqref{eq:berry_def} can be seen to be invariant modulo $2 \pi$.

There is also an explicitly gauge invariant expression for the Berry phase
\begin{gather} 
\phi(\gamma) = -\lim_{k\to \infty}\arg \prod\limits_{i=1}^{k} \langle \psi(x_i) | \psi(x_{i+1}) \rangle\nonumber\\
= -\lim_{k\to \infty} \arg P(x_1, x_2, \cdots, x_k),\label{eq:berry_inv}
\end{gather}
where $x_{i}$ are a $k$-point discretization of $\gamma$ that becomes infinitely fine as $k \to \infty$ and $x_{k+1} = x_{1}$.

Using Eq. \eqref{eq:berry_inv} and the reduction of an $n$-point function Eq. \eqref{eq:red} we can rewrite the Berry phase via 3-point functions

\begin{gather}\label{eq:berry_inv_int}
\phi(\gamma) = \lim_{k\to \infty} \sum_{i = 2}^{k-1} \Phi(x_1, x_i, x_{i+1})
=\int_{\gamma} A^{\gamma(0)},
\end{gather}
here $\gamma(0)$ is the origin point of the loop (which can be any point) and we have defined the one-form
\bea \label{eq:a_x_def}
\left. A^x(y) = A^x_{\alpha} dy^\alpha = dy^\alpha \partial_{z_\alpha} \Phi(x,y,z)\right|_{y=z}, 
\eea
that depends on $x$ as a parameter. Explicitly
\begin{gather*}
 A^x_{\alpha}(y) = i \left(\frac{\partial_{z_{\alpha}} P(x,y,z)|_{y=z}}{P(x,y)} - \frac{1}{2} \partial_{y_{\alpha}} \log P(x,y)\right).
\end{gather*}
As a matter of fact, the parameter $x$ of $A^x(y)$ in Eq. \eqref{eq:berry_inv_int} does not have to lie on the loop $\gamma$ and
\bea \label{eq:berry_int_any_x}
\phi(\gamma) = \int\limits_{\gamma} A^{x}
\eea
holds for any $x$. Comparing Eqs.\eqref{eq:berry_def} and \eqref{eq:berry_int_any_x} we conclude that $A^{x}(y)$ is the Berry connection in a specific gauge that depends on the reference point $x$.

To illustrate this, we show that if we change the reference point $X$ in the definition of $A^X_{\alpha}(y)$ the form changes by a gauge transformation. Indeed, let us consider two points $X_{1,2}$, then one finds
\begin{gather*}
    \Phi(X_2,y,z) = \Phi(X_1,y,z) + \Phi(X_2,X_1,y)+\Phi(X_2,X_1,z).
\end{gather*}
As the last step, using the definition of $A^{X_{1,2}}(y)$ we obtain
\begin{gather}
A^{X_2}_\alpha(y) 
= A^{X_1}_\alpha(y) + \partial_{y_\alpha} \Phi(X_2,X_1,y),
\end{gather}
with $\Phi(X_2,X_1,y)$ giving the desired gauge transformation.

One can further notice that the Berry curvature computed for $A^X$
\begin{gather}
    \partial_\alpha A^X_\beta(y) - \partial_\beta A^X_\alpha(y) = \omega_{\alpha\beta}(y)
\end{gather}
is independent of $X$ and is equal to the one computed for Eq. \eqref{eq:a_mu}.

For illustration, in Appendix \ref{app:spin_example} we computed $A^X(y)$ for a spin-$1/2$ particle in an external magnetic field. This case corresponds to Chern number $C=1$ and the Berry connection cannot be defined globally. In accordance with that, we see that $A^x(y)$ has a singularity when $y$ is the polar opposite of $x$.

What is remarkable, is that, unlike for the usual Berry phase $A$ 
the values of $A^x(y)$ are "gauge invariant". 
 Moreover, the phase of any three-point function can reconstructed from it:
\bea\label{eq:phase_recovery}
\Phi(x,y,z) = \int\limits_{\gamma_{xy}} \left(A^{z} - A^{x}\right),
\eea
where $\gamma_{xy}$ is any curve that connects points $x$ and $y$.

As we show in Appendix \ref{App:example}, there is an example of an immersion of a torus $T^2$ into $\mathbb{C}P^3$ where the Berry phase along any contractible loop is zero, but three-point functions have non trivial phases. This example proves that it is impossible to reconstruct phases of three point functions from Berry phases.

Finally, we find it useful to introduce the complexified version of the one-form defined in Eq. \eqref{eq:a_x_def}:
\bea
\mathcal{A}^{x}_{\alpha}(y) =   u_{\alpha} - i A_{\alpha}  = \label{eq:uglyA} 
 \partial_{z_{\alpha}}\left. \log P(x,y,z)\right|_{y=z}, 
\eea
where 
\bea
u_{\alpha}^{x}(y) = - \partial_{y_{\alpha}} \log\left(\cos d(x,y)\right).
\eea
If we integrate $u_{\alpha}^{x}(y)$ it is possible to reconstruct the quantum distance $D(x,y)$ between any two point.
Because of that and Eq. \eqref{eq:phase_recovery} we conclude that $\mathcal{A}^{x}_{\alpha}(y)$ contains all information about the 3-point function, and, thus, all geometry of the map $M \to \cpn$.
\section{Intrinsic geometry vs extrinsic geometry} \label{sec:int_vs_ext} 


The intrinsic geometry of a submanifold is loosely defined as all geometric information that can be measured within the submanifold itself without using the knowledge of how it is contained within the larger space. Any other information about the submanifold is considered extrinsic. This terminology is traditionally used for submanifolds of Riemannian manifolds, in which case all intrinsic geometry amounts to the knowledge of the induced metric \cite{Aminov2001}.

For the purposes of this work, we wish to define intrinsic geometry for submanifolds of spaces with both Riemannian and symplectic structure. We propose the following definition.

First of all, for the ambient space $L$ (with coordinates $y^{a}$) with the metric $g_{ab}$ and symplectic form $\omega_{ab}$, let us introduce a locally defined connection form $A_a$ that satisfies $d A = \omega$, or explicitly $\partial_a A_b - \partial_b A_a = \omega_{ab}$.

 Secondly, whenever a manifold $N$ (with coordinates $x^{\alpha}$) is mapped into $L$ a metric $g_{\alpha \beta}$ and one-form $A_{\alpha}$ are induced on $N$ in the following way (this operation is known as the "pullback"):
\begin{gather}
f: M \to L , \quad y^a = f^a(x^\alpha), \notag\\
g_{\alpha \beta} = \frac{\partial f^{a}}{\partial x^{\alpha}} \frac{\partial f^{b}}{\partial x^{\beta}} g_{ab},\quad
A_{\alpha} = \frac{\partial f^{a}}{\partial x^{\alpha}} A_{a}.
\end{gather}

These induced metric and one-form and everything that can be computed using them we define to be intrinsic geometry. And conversely, any property of the map $N \to L$ that does not reduce to the induced $g$ and $A$ is extrinsic.

The difference between intrinsic and extrinsic geometry is best illustrated by considering the distance between two points $x$ and $y$ in $N$. The intrinsic distance is defined as the length of the shortest path connecting $x$ and $y$
\begin{gather}
    d_N(x,y) = \min_{\gamma \subset N} \int\limits^1_0 \sqrt{g_{\alpha\beta}dx^\alpha dx^\beta},
\end{gather}
where the path $\gamma$ connects the points $x$ and $y$ within $N$. Similarly, if the shortest path is searched over all paths that connect the images of $x$ and $y$ in $L$, one obtains the extrinsic distance
\begin{gather}
    d_L(x,y) = \min_{\gamma \subset L} \int\limits^1_0 \sqrt{g_{a b}dx^a dx^b}.
\end{gather}
It is easy to see that $d_L(x,y) \leq d_N(x,y)$. However the expansions of both objects to the first two orders in $z = y - x$ are identical (see Appendix \ref{app:geodesics})
\begin{gather}
d^2_{N/L}(x,y) = g_{\alpha \beta} z^{\alpha} z^{\beta} +  \Gamma_{\alpha, \beta \gamma} z^{\alpha} z^{\beta} z^{\gamma} + O(z^4),  \label{eq:dist3rd}
\end{gather}
where
\bea \label{eq:christoffel}
\Gamma_{\alpha, \beta \gamma} = \frac{1}{2} (\partial_\beta g_{\alpha\gamma} + \partial_\gamma g_{\alpha \beta} -\partial_\alpha g_{\beta \gamma})
\eea
are the lower-index Christoffel symbols of $g_{\alpha \beta}$ and both $g$ and $\Gamma$ are evaluated at $x$. 
The right-hand side of Eq. \eqref{eq:dist3rd} depends on the intrinsic metric $g_{\alpha\beta}$ and its derivatives and, hence, is determined by intrinsic properties. The difference only appears in the next order
\begin{gather}\label{eq:N-L-diff}
d^2_{L}(x,y) - d^2_{N}(x,y)\\
= -\frac{1}{12}  \left(g^{ab}\Gamma_{a,\alpha\beta} \Gamma_{b,\gamma\delta} - g^{\epsilon\zeta}\Gamma_{\epsilon,\alpha\beta}\Gamma_{\zeta,\gamma\delta} \right) z^{\alpha} z^{\beta} z^{\gamma} z^{\delta} \notag,
\end{gather}
where $g^{ab}$ is the inverse of the metric in $L$ and $g^{\epsilon\mu}$ is the inverse of the submanifold metric. As we explain in Appendix \ref{app:geodesics}, this difference can be different for maps that share the same intrinsic geometry but have different extrinsic properties.
For detailed computations and useful concepts of extrinsic geometries of submanifolds, see Appendix \ref{app:extrinsicgeometry} and the textbook \cite{Aminov2001}.


In the context of this paper, $L$ is always the complex projective space $\cpn$. When considered as coming from a Hilbert space with a Hermitian product, $\cpn$ is endowed with a metric and a symplectic form, see Appendix \ref{sec:cpn}. In this case, the metric and the two form that are induced on $M$ are called the quantum metric and Berry curvature
\begin{gather}
\omega_{\alpha \beta}(x) = - 2 \im\langle\partial_{\alpha} \psi(x) | \partial_{\beta} \psi(x) \rangle\label{eq:def}\\
g_{\alpha \beta}(x) =  \re\left[\langle\partial_{\alpha} \psi(x) | \partial_{\beta} \psi(x) \rangle - \langle \partial_{\alpha} \psi | \psi \rangle \langle \psi | \partial_{\beta} \psi \rangle \right]\nonumber.
\end{gather}

As an example of a simple physical system, where the intrinsic and extrinsic distances are different, we can look at a spin-$1$ particle in a external magnetic field. The Hamiltonian reads as
\begin{gather*}
    H = \begin{pmatrix}
    B_z & \frac{1}{\sqrt{2}}\left(B_x + i B_y\right) & 0\\
    \frac{1}{\sqrt{2}}\left(B_x - i B_y\right) & 0  & \frac{1}{\sqrt{2}}\left(B_x + i B_y\right)\\
    0 & \frac{1}{\sqrt{2}}\left(B_x - i B_y\right) & -B_z
    \end{pmatrix}.
\end{gather*}
We parametrize the magnetic field as $B_z= B \cos \theta, B_x+i B_y = B \sin \theta e^{i\phi}$. Next, we consider the wave function corresponding to the largest positive eigenvalue $H\psi(\theta,\phi) = B\psi(\theta,\phi)$:
\begin{gather}
 \psi(\theta,\phi) =\left( e^{2 i \phi }\cos^2 \left(\frac{\theta}{2}\right),\frac{e^{i \phi }}{\sqrt{2}} \sin \theta, \sin^2 \left(\frac{\theta}{2}\right)\right).
\end{gather}
The quantum metric and Berry curvature Eq. \eqref{eq:def} can be seen to be 
\begin{gather}\label{eq:spin-1-state}
ds^2 =  \frac{1}{2}\left( d\theta^2 + \sin^2\theta d\phi^2\right),\quad \omega = - \sin \theta d\theta \wedge d\phi,
\end{gather}
which are just the metric and volume form of a 2-sphere (of radius $1/\sqrt{2}$). Accordingly, the intrinsic distance between any two states $L^{\rm int}$ is the spherical distance between the corresponding points. Meanwhile, the global distance computed from the overlap of the states Eq. \eqref{eq:spin-1-state} is
\begin{gather}
L^{\rm ext} = \arccos\left[\cos^2 \left(\frac{L^{\rm int}}{\sqrt{2}}\right) \right].
\end{gather}
One can verify that both distances agree for small $L_{int}$, but a deviation arises at the third order in agreement with Eq. \eqref{eq:N-L-diff}.

Extrinsic geometry also appears in the phase. If we consider the spherical triangle constructed on the points $A=(0,0), B=(\theta,0), C=(\theta,\phi)$ one finds for its area $S_{ABC} = \left(1 - \cos\theta\right) \phi$ while the phase of the corresponding three point function reads
\begin{gather}\label{eq:spin1-phase}
\alpha = \arg P^{(3)}(A,B,C), \quad \tan \frac{\alpha}{2} = \frac{\sin \phi}{\cot^2 \frac{\theta}{2}+\cos \phi}.
\end{gather}
The area of the triangle (that we compute by integrating $\omega$) is the intrinsic contribution to the phase, and the actual phase Eq. \eqref{eq:spin1-phase} agrees with it to the leading order in $\phi$ and $\theta$, but differs at higher orders where extrinsic contributions become important.

This example should be contrasted with the spin-$1/2$ case, where $L_{\text{int}} = L_{\text{ext}}$ and the three-point phase is equal to $S_{ABC}$.

Finally, we would like to note that along with quantitative differences in global phases and distances, extrinsic geometry can be responsible for discrete properties of immersions such as self-intersections that we discuss in Appendix \ref{app:intersection}.

In the next section we describe how $\omega$ and $g$ (as well as other objects) systematically arise in the expansion of $P(x,y,z)$ when $y,z$ are close to $x$.

\section{Local geometric objects} \label{sec:local}
We define a local geometric object as any gauge-invariant object composed out of $|\psi(x)\rangle$ and its derivatives at the same value of $x$. All such functionally independent objects have the from
\begin{gather}
c_{\alpha_1 \dots \alpha_{n}; \beta_1 \dots \beta_{m}}(x) = \notag\\
=\tr [P(x) \partial_{\alpha_1}\dots\partial_{\alpha_{n}} P(x) \partial_{\beta_1}\dots \partial_{\beta_{m}} P(x)],
\end{gather}
where the derivatives act on closest projector only.

However, since, according to Section \eqref{sec:berry_phase}, all information is contained in $\mathcal{A}^X(y)$, expanding $\mathcal{A}^{y+q}(y)$ in $q$ is sufficient to produce all independent structures. We parameterize this expansion as
\bea
\mathcal{A}_{\alpha}^{y + q}(y) = Q^{(1)}_{\alpha \beta}(y)q^{\beta} + Q^{(2)}_{\alpha \beta \gamma}(y) q^{\beta} q^{\gamma} + \cdots
\eea
From the calculation in Appendix \eqref{app:phase_geometry},
\begin{gather}
    Q^{(1)}_{\alpha\beta} = \tr[P(y)\partial_\alpha P(y) \partial_\beta P(y)] =  g_{\alpha\beta} - \frac{i}{2} \omega_{\alpha\beta}, \notag
\end{gather}
is recognized as the quantum geometric tensor. The next structure
\bea
Q^{(2)}_{\alpha\beta\gamma} = \tr\left[P(x)\partial_\alpha P(x) \partial_{\beta} \partial_{\gamma}P(x)\right] \label{eq:defq2q3}
\eea
is the quantum geometric connection introduced in \cite{Ahn_2020, Kozii_2021},
that can be decomposed as 
\begin{gather}
Q^{(2)}_{\alpha\beta\gamma} =\Gamma_{\alpha, \beta \gamma}(x) - \frac{i}{2} \tilde{\Gamma}_{\alpha, \beta \gamma}(x), \label{eq:cabg}
\end{gather}
where the real part $\Gamma_{\alpha, \beta \gamma}(x)$ are the lower index Christoffel symbols from Eq. \eqref{eq:christoffel} which only contain intrinsic information and $\tilde{\Gamma}_{\alpha, \beta \gamma}(x) = \omega_{\alpha a} \Gamma^{a}_{\beta \gamma}$ which, since one contraction is performed over a $\cpn$ index $a$, is sensitive to extrinsic geometry.

For non-degenerate $\omega_{\alpha\beta}$, $\tilde{\Gamma}_{\alpha, \beta \gamma}(x)$ enjoys geometric interpretation as a {\it symplectic connection} \footnote{The general definition of connections (or, equivalely, covariant derivatives) can be found in \cite{nakahara2018geometry} and symplectic connections are introduced in \cite{fedosov1996deformation}.}.  The reasoning goes as follows. Using the definition Eq. \eqref{eq:defq2q3}, one can derive 
the following identity
\bea \label{eq:omega_conservation}
\partial_\alpha \omega_{\beta\gamma} = - \tilde{\Gamma}_{\gamma,\beta\alpha} + \tilde{\Gamma}_{\beta,\gamma\alpha}.
\eea
Since $\omega_{\alpha\beta}$ is non-degenerate we can define a connection $\tilde{\nabla}_{\alpha}$ with components 
$\tilde{\Gamma}^{\alpha}_{\beta \gamma}(x) = \omega^{\alpha \tau}\tilde{\Gamma}_{\tau, \beta \gamma}(x)$, and Eq. \eqref{eq:omega_conservation} reads as
\bea \label{eq:omega_conserv}
\tilde{\nabla}_{\alpha}\omega_{\beta \gamma} = 0.
\eea
Eq. \eqref{eq:omega_conserv} means that $\omega_{\alpha \beta}$ is preserved under parallel transport with $\tilde{\nabla}_{\alpha}$ which is precisely the definition of a symplectic connection.

It is a standard result \cite{fedosov1996deformation} on symplectic connections, that for a given $\omega$ Eq. \eqref{eq:omega_conservation} does not determine $\tilde{\Gamma}_{\gamma,\beta\alpha}$ uniquely.
More specifically, for a given $\omega$ any symplectic connection can be written as
\bea \label{eq:t_def}
\tilde{\Gamma}_{\gamma, \alpha \beta} = \omega_{\gamma \delta}\tilde{\Gamma}^{\delta}_{\alpha \beta } = \frac{1}{3}(\partial_\beta \omega_{\gamma \alpha} + \partial_\alpha \omega_{\gamma \beta}) + T_{\alpha \beta \gamma},
\eea
where $T_{\alpha \beta \gamma}$ is some fully symmetric rank-3 tensor and, conversely, for every fully symmetric rank-3 tensor $T_{\alpha \beta \gamma}$ Eq. \eqref{eq:t_def} is a symplectic connection. Note that, the definition of $T_{\alpha \beta \gamma}$ Eq. \eqref{eq:t_def} does not require $\omega$ to be non-degenerate.

In \cite{Ahn_2020} it was shown that when the map is from a 2-torus $\mathbb{T}^2$ to $\mathbb{C}P^1$, the following identity holds:
\begin{gather*}
\tilde{\Gamma}^{\alpha}_{ \beta \gamma} = \Gamma^{\alpha}_{ \beta\gamma},
\end{gather*}
showing that, in this particular case, $T_{\alpha \beta \gamma}$ is completely determined by $g$ and $\omega$. 
This can be explained by the fact that $T^2$ and $\mathbb{C}P^1$ have the same dimension and therefore extrinsic geometry does not play a role. In general,  $\tilde{\Gamma}_{\alpha,\beta\gamma}$ is not uniquely determined by $g$ and $\omega$.

In particular, Appendix \ref{App:example} provides an example of an immersion $\mathbb{T}^2$ to $\mathbb{C}P^3$ where $g$ is flat, $\omega = 0$, but $T_{\alpha \beta \gamma}$ can vary. This establishes $T_{\alpha\beta\gamma}$ as an independent gauge invariant object that describes the geometry of a set of states along with $g$ and $\omega$. Where we decided to use $T_{\alpha \beta \gamma}$ to describe the new geometry because of its greater symmetry and since $T_{\alpha \beta \gamma}$ and $\tilde{\Gamma}_{\gamma, \alpha \beta}$ encode the same information (assuming $\omega$ is known).

Furthermore, in general, even the knowledge of $g$, $\omega$ and $T$ does not fix the map $M \to \cpn$ completely, but there are special cases when it does. For example, in Appendix \ref{app:reconstruct}, we prove that maps $S^1 \to \mathbb{C}P^1$ can be uniquely reconstructed from $g$ and $T$ ($\omega$ is identically zero for one dimensional manifolds).



\section{Holomorphic immersions, Landau levels, tBLG} \label{sec:holomorphic}
In this section we will describe the additional special properties of geometry when $M$ is a two-dimensional complex manifold (we will call it $M_2$ from now on) and the map $f: M_2 \to \cpn$ is holomorphic. This means that we can introduce a complex coordinate $y\in \mathbb{C}$ on $M_2$ and $n$ complex coordinates $z_i$ on $\cpn$ such that
\begin{gather}
   z_i = f_i(y),\quad  \bar{\partial}_{\bar{y}} f_i = \partial_y \bar{f}_i = 0,
\end{gather}
in this case the metric and Berry curvature could be expressed in a quite simple form
\begin{gather}
    g_{y\bar{y}} = \partial_y \bar{\partial}_{\bar{y}} \log K(y,\bar{y}), \quad g_{yy}=g_{\bar{y}\bar{y}} = 0, \notag\\
    \omega_{y\bar{y}} = - \omega_{\bar{y}y} = i  \partial_y \bar{\partial}_{\bar{y}} \log K(y,\bar{y}), \notag\\
    K(y,\bar{y}) = 1+ \sum^n_{i=1} \left|f_i(y)\right|^2, 
\end{gather}
which means that intrinsic properties of $M_2$ are completely determined by a single function $K(y,\bar{y})$. What is more remarkable, is that $K(y,\bar{y})$ also completely determines the functions $f_i(y)$. Namely, if we have two sets of functions $f_i(y)$ and $g_i(y)$ such that
\begin{gather}
    \sum^n_{i=1} \left|f_i(y)\right|^2 = \sum^n_{i=1} \left|g_i(y)\right|^2,
\end{gather}
one can find a unitary matrix $M_{ij}$ such that $f_i(y) = M_{ij} g_j(y)$. This is known as the {\it Calabi rigidity} \cite{calrig}. It implies that, in principle, if one knows the function $K(y,\bar{y})$ the map $f: M_2 \to \cpn$ can be uniquely restored up to a holomorphic isometry of $\cpn$. Moreover, if we know just the Berry curvature we can restore the function $K(y,\bar{y})$ and therefore in the case of a holomorphic immersion all observables are expressible in terms of the Berry curvature and its derivatives only. 

We now show how this general fact can be helpful for understanding of the properties of Landau levels and the chiral limit of twisted bilayer graphene (see Appendices \ref{sec:GLL} and  \ref{sec:TBG} as well as \cite{popov2021hidden,tarnopolsky2019origin} for more detail). In both cases, the Bloch functions are holomorphic functions of the quasi-momentum
\begin{gather}
    \psi^\alpha_{k}(x,y) \propto u_k(z=x+i y) \psi^\alpha_{0}(x,y),
\end{gather}
where the function $u_k$ is defined in \eqref{eq:holfun} and is the same in both cases, but $\psi_0^{\alpha}$ is different. Therefore, the map $f$ to $\mathbb{C}P^\infty$ is holomorphic and the Calabi rigidity applies. Which, in turn, implies that all extrinsic properties are completely determined by intrinsic geometry.

For instance, for the lowest Landau Level the quantum metric and Berry curvature are computed to be
\bea \label{eq:Landau_geometry}
    g_{ij} = \frac{1}{B} \delta_{ij},\quad \omega_{ij} = \frac{1}{B}\epsilon_{ij},
\eea
where $\delta_{ij}$ is the Kronecker delta and $\epsilon_{ij}$ is the Levi-Civita tensor. Meanwhile, the argument of three-point function 
\begin{gather}
    \arg P^{(3)}(q,k,p) \propto \operatorname{Area}(q,k,p) \label{eq:areaLL}
\end{gather}
is proportional to the area of the triangle constructed on the vectors $q,k,p$ 
and the quantum distance is
\begin{gather}
    D^2(k, p) = 1 - \exp\left(-\frac{\left|k-p\right|^2}{2B}\right). \label{eq:quantdistLL}
\end{gather}
In this case, the Calabi rigidity states that any holomorphic immersion into $\cpn$ such that metric and Berry curvature are given by Eq. \eqref{eq:Landau_geometry} also has the three-point function with the phase Eq. \eqref{eq:areaLL} and quantum distance Eq. \eqref{eq:quantdistLL}.

\section{Band structure calculations} \label{sec:bands}

The formalism developed so far has an immediate application to any calculation involving a band structure. When we consider a $d$-dimensional crystalline system, according to the Bloch theorem, the eigenfunctions have the form $|\psi_m(k)\rangle = e^{i k \hat{r}} |u_m(k) \rangle$, where $k$ is the quasimomentum, $m$ is the band index, $\hat{r}$ is the position operator and $|u_m(k) \rangle$ is periodic with respect to crystal translations. The quasimomentum $k$ is a $d$-dimensional vector that is defined up to a dual lattice translation and, thus, forms the Brilloiun zone (BZ) that is topologically equivalent to a $d$-torus $\mathbb{T}^d$. Let us also assume that there are $n+1$ orbitals within each unit cell. With that, each Bloch function $|u_m(k) \rangle$ can be thought of as defining a map $P_m(k): \mathbb{T}^d \to \cpn$.

\subsection{Relation to polarization distribution}

Below we show how the objects introduced in the Modern theory of polarization  \cite{Resta2007, PhysRevB.47.1651,RevModPhys.66.899} are naturally rewritten in terms of the geometric objects of this paper.

The observable we wish to describe is the polarization density of a sample

\begin{gather}
\hat{P} = \frac{e \hat{X}}{V},\quad  \hat{X} = \sum_i \hat{x}_i, 
\end{gather}
where $\hat{x}_i$ are the position operators for all electrons and we have ignored the ionic contribution. There are many subtleties in defining polarization as a bulk property, for a detailed discussion, please, see \cite{Resta2007}. The most important formal result is that bulk polarization is only defined up to a polarization quantum.

As was established in \cite{PhysRevB.47.1651}, for a band insulator with the first $n$ bands filled the average polarization is given by
\bea \label{eq:ave_P_def}
\langle \vec{P} \rangle = e \sum\limits_{m \in \text{occ}} \vec{A}_m =  e\sum\limits_{m \in \text{occ}}\int
\frac{d^d k}{(2\pi)^d} \vec{A}_{m}(k),
\eea
where $\vec{A}_{m}(k)$ is the Berry connection of the $m$-th band and $\vec{A}_{m}$ is $\vec{A}_{m}(k)$ averaged over the Brillouin zone. In accordance with ambiguity of $\hat{P}$, the right hand side of Eq. \eqref{eq:ave_P_def} is also only gauge invariant up to a shift by the polarization quantum.

Eq. \eqref{eq:ave_P_def} was later extended to the whole polarization distribution and not just its average \cite{SouzaWilkensMartin2000} by considering the generating function for the cumulants of $\hat{X}$:
\bea
C(q) = \langle e^{-i q \hat{X}} \rangle,
\eea
where the average is taken over the many-body wavefunction. When only one band with the wavefunction $| u_{k} \rangle$ is occupied (we confine ourselves to this case only), we obtain a compact expression
\bea
\frac{\log C(q)}{V} = \int \frac{d^d k}{(2\pi)^d} \log \braket{ u_k | u_{k + q}}.
\eea
Remarkably, the generating function has an equally simple expressions in terms of gauge invaraint objects (see Appendix \ref{app:polarization} for derivation): 
\begin{gather}
\frac{\log C(q)}{V} =  \vec{q} \cdot \vec{A}
+ q^\beta \int \frac{d^d k}{(2\pi)^d} \int_{0}^{1} dt \left[ \bar{\mathcal{A}}^{k}_{\beta}(k + t q) \notag \right].
\end{gather}
After expanding $\mathcal{A}$, this leads to
\begin{gather*}
\frac{\log C(q)}{V} = \vec{q} \cdot \vec{A}
+ \sum_{n}\int \frac{d^d k }{(2\pi)^d}   \frac{\bar{Q}^{(n)}_{\alpha_1 \dots \alpha_{n+1}}(k)}{n+1}  q^{\alpha_1} \cdots q^{\alpha_{n+1}},
\end{gather*}
which allows us to write for the cumulants ($ n \geq 2$)
\bea
\langle X_{\alpha_1} \cdots X_{\alpha_n} \rangle_c = V \int  \frac{d^dk}{(2\pi)^d}  \bar{Q}^{(n)}_{(\alpha_1 \dots \alpha_{n+1})}(k),
\eea
where the brackets $(,)$  denote symmetrization and $c$ in the subscript of $\langle \rangle_c$ means that we work with cumulants and not moments.
In particular, 
\begin{gather*}
\langle X_{\alpha} X_{\beta} \rangle_c =  V \int \frac{d^d k}{(2\pi)^d}  g_{\alpha \beta}(k),\notag\\
\langle X_{\alpha} X_{\beta} X_{\gamma} \rangle_c = V \int \frac{d^d k}{(2\pi)^d}  T_{\alpha \beta \gamma}(k).
\end{gather*}

We note, that the expression for $\langle X_{\alpha} X_{\beta} X_{\gamma} \rangle_c$ in two-band models was previously reported in \cite{Kozii_2021} and \cite{Patankar_2018}.

\subsection{Conductivity in an inhomogenous electric field} 
As the next example, we look at the calculation of conductivity of a free electron material in a modulated electric field $\vec{E} = \vec{E}_0 e^{i \vec{q}\cdot\vec{r}}$. We will closely follow the exposition in \cite{Kozii_2021}.
The response has the form:
\bea
j_{\alpha}(q) = \sigma_{\alpha\beta}(q) E_{0}^{\beta}(q),
\eea
where $\sigma_{\alpha \beta}(q)$ is the conductivity tensor to be computed.

Previously, conductivity in nonuniform fields has been studied in \cite{Kozii_2021, Zhang_2022}, but these calculations focused on two-dimensional two-band models where only intrinsic geometry is present. Additionally only expansions up to the second order in $q$ were computed.

Here we present a calculation of the conductivity for a system with an arbitrary number of bands and at finite $q$. However, in order to obtain an answer where the response is determined solely by geometry we consider the limit where only the one lowest band is filled. We further assume that this band is separated from all other bands by a gap $\Delta$, all bands are flat and all unoccupied bands are at the same energy. An example of a non-trivial model that satisfies these conditions is given in the next subsection.

If we define the projector on the single occupied band as $P(k) = |u_0(k) \rangle \langle u_0(k) |$ and $\mathcal{A}^X(y)$ is defined according to Eq. \eqref{eq:uglyA} the conductivity reads 
\begin{widetext}
\begin{gather*} \label{eq:general_sigma}
\sigma_{\alpha\beta}(q) = i\int \frac{d^dk}{(2\pi)^d} \left[P(k,k+q/2) \left(\mathcal{A}^{k + \frac{q}{2}}_{\alpha}(k) \bar{\mathcal{A}}^{k+ \frac{q}{2}}_{\beta}(k) + \bar{Q}_{\alpha \beta}(k)\right)-P(k,k-q/2) \left(\bar{\mathcal{A}}^{k -  \frac{q}{2}}_{\alpha}\left(k \right) \mathcal{A}^{k - \frac{q}{2}}_{\beta}\left(k\right) + Q_{\alpha \beta}(k) \right) \right],  
\end{gather*}
\end{widetext}
with the details of the calculation delegated to Appendix \ref{app:conduct}.
In order to compare with the results in \cite{Kozii_2021}, we expand this expression to the second order in $q$ and also choose $q^{\alpha} = q \delta^{\alpha x}$, obtaining
\bea
\sigma_{xy} = - \int \frac{d^dk}{(2\pi)^d} \omega_{xy} \left( 1 - g_{xx} \frac{q^2}{2} \right)
\eea
which agrees with \cite{Kozii_2021} in the flat band limit.

\subsection{Veronese band structure}

As an example of a model where the Berry curvature and quantum metric are trivial, but the geometry as a whole is not, we consider the following Hamiltonian (see Appendix \eqref{App:example})
\begin{gather} \label{eq:local_veronese}
h(k) = \mathbf{1}_{4} - |\psi(k) \rangle \langle \psi(k) |,\\
|\psi(k) \rangle = \{e^{i (k_x n + k_y m)} c_x c_y, 
 e^{i k_y m} c_y s_x, e^{i k_x n} c_x s_y, 
 s_x s_y\}\nonumber,
\end{gather}
here $\mathbf{1}_{4}$ is the 4 by 4 identity matrix, $c_{x/y} = \cos k_{x/y}$, $s_{x/y} = \sin k_{x/y}$ and $n,m$ are integer numbers. This model model corresponds to a local hopping Hamiltonian in real space because Eq. \eqref{eq:local_veronese} is a trigonometric polynomial in $k_{x/y}$.

This Hamiltonian describes 4 perfectly flat bands. The bottom band is non-degenerate at zero energy and the top three bands have equal energies of $1$. The quantum metric and Berry curvature of the bottom band are
\begin{gather*}
g = \text{diag}\left(\frac{1}{1 + n^2 \sin^2(k_x)}, \frac{1}{1 + m^2 \sin^2(k_y)}\right),\\
\omega = 0.
\end{gather*}
Note that by reparameterizing $k_x$ and $k_y$ the metric can be made flat $g = \text{diag}(1,1)$. The 3-tensor $T$ reads
\begin{gather*}
T_{xxx} = \frac{n \cos(k_x)}{2} (2 + n^2 \sin^2(k_x)),\\
T_{yyy} = \frac{m \cos(k_y)}{2} (2 + m^2 \sin^2(k_y))
\end{gather*}
and all other components are zero. The fact that $T$ is non-trivial  shows that, while Berry phases for all contractible loops vanish for this band structure, 3-point functions support non-zero phases. 

When only the bottom band is filled we expect all low-energy properties to be functions of just the geometry of the this band and the leading behavior of the phenomena sensitive to the phase of Bloch wavefunctions to be determined by $T$. 

\section{Discussion and Outlook} \label{sec:disuccsion}


In this paper we developed a framework for describing all gauge invariant properties of a submanifold of quantum states via the three point function $P^{(3)}(x,y,z)$. 
We showed that the knowledge of these functions uniquely determines the immersion of the submanifold into the complex projective space $\cpn$ up to the action of a holomorphic isometry of $\cpn$.  Being a complete description, our approach allows one to view all previously studied geometric objects in a unified way. Moreover, certain objects that appear in the expansion of $P^{(3)}$, in particular the 3-tensor $T_{\alpha \beta \gamma},$ are sensitive to extrinsic geometry, and, as a consequence, are functionally independent from intrinsic objects such as the quantum metric and Berry curvature. As an illustration for that, we constructed an immersion where the metric is flat and the Berry curvature vanishes but the three tensor $T_{\alpha\beta\gamma}$ can vary. This shows, that one should not rely on the quantum geometric tensor as the complete description of a manifold quantum states, but consider the extrinsic geometry as well, for example, by working with $\mathcal{A}^X(y)$. In contradistinction to the general case, we prove that when the parameter space is 2 dimensional and the map to $\cpn$ is holomorphic, the quantum geometric tensor is sufficient to fully reconstruct the map.

An important result of our work is the procedure that employs the interpretation of $P^{(3)}$ in terms of geodesic triangles in $\cpn$ to relate the terms in the local expansion of $P^{(3)}$ to geometric objects in an efficient way.  

We envision our approach to find most uses in condensed matter calculations involving electronic band structures. In Section \eqref{sec:bands} we have shown how the distribution of polarization in insulators is straightforwardly rewritten in terms of $\mathcal{A}^{X}(y)$ and the moments of polarization distribution are expressed in terms of the objects $Q^{(n)}$ that appear in the expansion of $\mathcal{A}^{X}(y)$. This can potentially lead to a direct experiment measurement of $T_{\alpha \beta \gamma}$. Likewise, we were able to express the conductivity of a flat band material  in a modulated field $\sigma_{\alpha \beta}(q)$ in terms of $\mathcal{A}^{X}(y)$. Expanding $\sigma_{\alpha \beta}(q)$ in $q$ again leads to direct ways of measuring local structures $Q^{(n)}_{\alpha_1 \dots \alpha_{n+1}}(k)$.

As the next step in the development of our approach, we would like to understand how our analysis generalizes to the case of projectors on  $k$-dimensional subspaces (here only a single state projector has been considered). In this case, one would need to study maps to the complex Grassmannian $Gr(n + 1,k)$.

Another direction where progress can be made is the refinement of our approach to the case where the map to $\cpn$ satisfies certain constraints. These could be coming from discrete symmetries in solid state systems such as time reversal and inversion and well as the specifics of the crystalline group.

We would like to particularly stress possible applications to correlated flat-band physics where band dispersion vanishes and the dynamics are governed by geometry. An interesting example is the geometric condition for the formation of a Fractional Chern insulator from flat bands\cite{Roy_2014} and the superfluid weight of a flat-band superconductor\cite{Peotta_2015} that has been formulated using $\omega$ and $g$ . One might wonder if these results can be improved if the full characterization of the geometry is used.



\section{Acknowledgments}
We are grateful to J. E. Moore, D.S.Antonenko, V.Kozii, C. 
Liu, D. Chowdhury and D. Mao for useful discussions. F.K.P. is currently a Simons Junior Fellow at NYU and supported by a grant 855325FP from the Simons Foundation. A. A. acknowledges support from the NSF under grant number DMR-1918065 and from a Kavli ENSI fellowship.

\bibliography{geometry}

\newpage

\appendix
\begin{widetext}

\begin{center}
\textbf{\large Supplemental Materials for ``Extrinsic geometry of quantum states''} 
\end{center}


\section{Geometry of complex projective spaces}
\label{sec:cpn}
In this section we will use notation and concepts from complex geometry a good review of which can be found in \cite{griffiths2014principles}.
The complex projective space, $\cpn$, is defined as a space of all possible directions in the linear complex space $\mathbb{C}^{n+1}$. To be exact, $\cpn$ is defined to be the set of the equivalence classes of the  following equivalence relation on $\mathbb{C}^{n+1}$
\begin{gather}
   \left(z_1:z_2:\ldots:z_{n+1}\right) \sim \lambda  \left(z_1:z_2:\ldots:z_{n+1}\right),
\end{gather}
where $\lambda$ is any non-zero complex number.  Thus, any point in $\cpn$ is represented by a vector in $\mathbb{C}^{n+1}$ modulo multiplication by a non-zero complex number. Let us consider the subset of vectors with $z_{n+1} \neq 0$. In this case, we can set $z_{n+1}=1$ and use the other complex coordinates $\left\{z_i\right\}^{n}_{i=1}$  to parameterize almost all points in $\cpn$. Using the definitions Eqs. \eqref{eq:def}, we find for the 2-from and metric
\begin{gather}\label{eq:cpn_o_g}
    \omega = \omega_{i\bar{j}}dz_i \wedge d\bar{z}_j =  - i \frac{dz_i \wedge d\bar{z}_i}{1 + \bar{z}_k z_k} + i\frac{\bar{z}_i z_j dz_i \wedge d\bar{z}_j}{(1+\bar{z}_k z_k)^2} , \quad ds^2 = g_{i\bar{j}} dz_i d\bar{z}_j =  \frac{dz_i d\bar{z}_i}{1+\bar{z}_k z_k} - \frac{\bar{z}_i z_j dz_i d\bar{z}_j}{(1+\bar{z}_k z_k)^2}.
\end{gather}
One can also introduce a complex structure by
\begin{gather}
   J(dz^j) = i dz^j, \quad J(d\bar{z}^j) = - i d\bar{z}^j.
\end{gather}
It can be verified that thus defined the complex structure is "covariantly constant": $\nabla J = 0$, where $\nabla$ is the covariant derivative constructed from the metric in Eq. \eqref{eq:cpn_o_g}. The same is true for the two-form
\begin{gather}
    g(u,v) = \omega(u,Jv) \Rightarrow \nabla \omega = 0.
\end{gather}
If we have two points in $\cpn$ represented by two unit vectors $\ket{x}$ and $\ket{y}$ in $\cpn$,  we can find that the geodesic connecting these two points in $\mathbb{C}^{n+1}$ is represented by the following line
\begin{gather}
    \ket{z(t)} = t \ket{x} + (1-t) e^{-i \beta} \ket{y},\quad t \in \left[0,1\right], \quad  \braket{x|y} = e^{i\beta} \left|\braket{x|y}\right| 
\end{gather}
The geodesic distance between any two points $|x\rangle$ and $|y\rangle$ is computed to be
\begin{gather}
    d(x,y) =  \arccos \left[ \frac{\left|\braket{x|y}\right|}{\sqrt{\braket{x|x}\braket{y|y}}}\right].
\end{gather}

\section{Complex projective GPS theorem or Quantum Tomography} \label{app:gps_theorem}


Let us show that if we have a set of $N\geq n+1$ points of general position in $\cpn$, we can restore theirs positions uniquely up to the action of a {\it holomorphic isometry} of $\cpn$ (a transformation that preserves both the distance and complex structure) if we know all three- and two-point functions $P^{(2)}(x_i,x_j)$ and $P^{(3)}(x_i,x_j,x_k)$. We briefly note, that if we only knew $P^{(2)}(x_i,x_j)$ we could uniquely determine the positions only up to the action of all isometries of $\cpn$ as a real manifold, which encompasses all transformations that preserve the metric but not necessarily the complex structure.  

The proof is simple. Consider two sets of points $x_i$ and $y_j$ such that $P^{(3)}(x_i,x_j,x_k) = P^{(3)}(y_i,y_j,y_k)$ for all $i,j,k$. Now we represent each of the points $x_i$ and $y_i$ by some unit vector in $\mathbb{C}^{n+1}$ $\ket{x_i} e^{i\alpha_i}$ or $\ket{y_i} e^{i\beta_i}$ defined up to phases $\alpha_i,\beta_i$. In what follows, it suffices to set $\alpha_i=0$. Let us look at the first $n+1$ points from each set. Because of the generality condition they form bases in $\mathbb{C}^{n+1}$. We compute the coefficients of the Hermitian form in these bases
\begin{gather}
    h^x_{ij} = h(x_i,x_j) = \braket{x_i| x_j} = r_{ij} e^{i\phi_{ij}}, \quad h^y_{ij} = h(y_i,y_j) = \braket{y_i| y_j} e^{i\left(\beta_i - \beta_j\right)} = r_{ij} e^{i\left(\varphi_{ij} + \beta_i - \beta_j\right)},
\end{gather}
where we have used that $\left|\braket{y_i|y_j}\right| =\left| \braket{x_i|x_j}\right|$. The phases $\varphi_{ij}$ and $\phi_{ij}$ satisfy the cocycle condition $\varphi_{ij} + \varphi_{jk} + \varphi_{ki} = \phi_{ij} + \phi_{jk} + \phi_{ki}$. Using this equation we can always find $\beta_i$ such that $\phi_{ij} = \varphi_{ij} + \beta_i-\beta_j$.
For instance, we can use $\beta_i=\phi_{i1}-\varphi_{i1}$. Now, we define $\ket{\bar{y}_i} = e^{i\beta_i} \ket{y_i}$. With that,
\begin{gather}
    \braket{x_i|x_j} = \braket{\bar{y}_i|\bar{y}_j}, \text{implying that we can find a unique } g \in SU(N+1), \text{such that}\quad g\ket{\bar{y}_i} = \ket{x_i}
\end{gather}
which, in turn, means that with the use of a $SU(N+1)$ transformation we can map $y_i$ to $x_i$. This proves the statement for $N=n+1$, because there is a one-to-one correspondence between elements of $SU(N+1)$ and holomorphic isometries of $\cpn$.
As the next step, we show that the relation $g\ket{y_k} = \ket{x_k}$ also holds for $k>n+1$. Indeed, since first $n+1$ elements in both sets form bases we have
\begin{gather}
    \ket{y_k} =\sum^{n+1}_{i=1} Y^k_i \ket{\bar{y}_i}, \quad  \ket{x_k} =\sum^{n+1}_{i=1} X^k_i \ket{x_i},
\end{gather}
and one can see that $\left|Y^k_i\right| = \left|X^k_i\right|$ because $P^{(2)}(y_k,y_i) = P^{(2)}(x_k,x_i)$. But since $P^{(3)}(y_k,y_i,y_j) = P^{(3)}(x_k,x_i,x_k)$ further implies $\braket{\bar{y}_i|\bar{y}_j} = \braket{x_i|x_j}$ and we get
\begin{gather}
    \braket{y_k|\bar{y}_i} \braket{\bar{y}_j|y_k} = \braket{x_k|x_i} \braket{x_j|x_k} \Rightarrow  \frac{\braket{x_k|x_i}}{\braket{y_k|\bar{y}_i}} = \frac{\braket{\bar{y}_j| y_k}}{\braket{x_j|x_k}},
\end{gather}
these equation are only satisfied for all $i$ and $j$ if and only if there exists a global phase $e^{i\gamma}$ such that $Y^k_i = X^k_i e^{i\gamma}$. With that we redefine $\ket{\bar{y}_k} = e^{i\gamma} \ket{y_k}$, leading to
\begin{gather}
    \ket{\bar{y}_k} = \sum\limits^{n+1}_{i=1} X^k_i \ket{\bar{y_i}}, \quad g\ket{\bar{y}_k} = \sum\limits^{n+1}_{i=1} X^k_i g\ket{\bar{y_i}} = \sum\limits^{n+1}_{i=1} X^k_i \ket{\bar{x_i}} = \ket{x_k}.
\end{gather}
Therefore, we conclude that the constructed $g \in SU(N+1)$ maps all $\ket{y_i}$ to $\ket{x_i}$ and provides the desired holomorphic isometry of $\cpn$.

\section{$A^{X}(y)$ for a spin in an external magnetic field} \label{app:spin_example}
In this section we will compute the the generalized connection form $A^{X}(y)$ for the simple system of a spin-$1/2$ particle in an external magnetic field. The Hamiltonian is
\bea
H(\boldsymbol{B}) = \boldsymbol{B} \cdot \boldsymbol{\sigma}, \quad \boldsymbol{B} = \left(B_x,B_y,B_z \right),\quad B = \sqrt{\boldsymbol{B}^2}, \quad \boldsymbol{\sigma} = \left(\sigma_x,\sigma_y,\sigma_z\right) 
\eea
where $\sigma_x,\sigma_y,\sigma_z$ are the Pauli matrices. The ground state of this Hamiltonian is 
\bea
| \boldsymbol{B} \rangle = \left\{ \frac{B_z - B}{B_x + i B_y}, 1 \right\} \sim \left\{1, -\zeta \right\},
\eea
where we have introduced a complex coordinate $\zeta = x + i y$. The phase of the three point function for the ground state is
\bea
\Phi(\zeta_1, \zeta_2, \zeta_3) = i\log\frac{(1 + \bar{\zeta}_1 \zeta_2)(1 + \bar{\zeta}_2 \zeta_3)(1 + \bar{\zeta}_3 \zeta_1)}{\left|(1 + \bar{\zeta}_1 \zeta_2)(1 + \bar{\zeta}_2 \zeta_3)(1 + \bar{\zeta}_3 \zeta_1)\right|}.
\eea
Let us now compute the one from. First, we set $\zeta_3 = 0$, in which case,
\begin{gather}
    \Phi(\zeta_1,\zeta_2,0) = i\log\left[ \frac{1+ \bar{\zeta}_1 \zeta_2}{|1+ \bar{\zeta}_1 \zeta_2| }\right], \quad A^0(x+ i y) = A^{0}(\zeta=x+ i y) = \frac{y dx -  x dy }{1 + x^2 + y^2}
\end{gather}
from that it can be seen that the one-form has a singularity at infinity $\zeta  =\infty$. To show it explicitly, we look at the one-form for an arbitrary reference point $\zeta = x_1 + i y_1$



\bea \label{eq:general_ref_A}
\boldsymbol{A}^{\boldsymbol{r}_1}(\boldsymbol{r}_2) = - \frac{\boldsymbol{\epsilon} \cdot \boldsymbol{r}_2}{1 + r_2^2} + \frac{\boldsymbol{\epsilon} \cdot  (\boldsymbol{r}_1 + \boldsymbol{r}_1^2 \boldsymbol{r}_2) }{1 + 2 \boldsymbol{r}_1 \cdot \boldsymbol{r}_1 + \boldsymbol{r}_1^2 \boldsymbol{r}_2^2 },
\eea
where bold symbols were used vectors and the Levi-Civita tensor $\boldsymbol{\epsilon}$, $\cdot$ stands for matrix-vector multiplication and $\boldsymbol{r}_i = \{x_i, y_i\}$. Eq. \eqref{eq:general_ref_A} is regular at the infinity $r^2 = x^2 + y^2 \to \infty$, but has a simple pole at $x_p + i y_p = \frac{1}{\zeta}$.

\section{Extrinsic Geometry} \label{app:extrinsicgeometry}
In this section we give a short review of the subject of extrinsic geometry of Riemannian submanifolds. For more information we refer the reader to classic textbook \cite{Aminov2001}. A submanifold is defined via a map $f:N^m \to M^n$, where we assume that $M^n$ is a Riemannian manifold with a metric $g_{a b}$. One can look at this map as specifying $n$ functions of $m$ variables
\begin{gather}
    x^{a}(\xi_1,\ldots,\xi_m).
\end{gather}
Now we define a tangent vector and a set of normal vectors at each point of our original manifold:
\begin{gather}
    e_\alpha^{a}(\xi_1,\ldots,\xi_n) =  \partial_\alpha x^{a}, \quad  g_{a b }n_i^{b} e_\alpha^{a} = 0, \quad g_{a b} n_i^{a} n_j^{b} = \delta_{ij}.
\end{gather}
With that, the induced metric on $N$ is defined as $h_{\alpha\beta} = e_\alpha^{a} e_\beta^{b} g_{a b}$.

Next, we compute covariant derivatives in the direction of each tangent vector $e_\alpha^{a}$, $\nabla_\alpha = e^{a}_\alpha \nabla_{a}$: 
\begin{gather}
    \nabla_\alpha e^{a}_\beta = \Gamma^\gamma_{\alpha\beta} e^{a}_\gamma + K^i_{\alpha\beta} n_i^{a}, \quad \nabla_\alpha n^{a}_i = - K^i_{\alpha\beta} h^{\beta\gamma} e^{a}_{\gamma} + A_{\alpha, ij} n_j^{a}.
\end{gather}
here $\Gamma^\gamma_{\alpha\beta}$ is an induced Christoffel symbols on the submanifolds, the symmetric form $K^i_{\alpha\beta}$ is the second fundamental form or shape tensor and the antisymmetric-matrix-valued one-form $A_{\alpha, ij}$ is the Gauss torsion or normal connection. 
Along with that, we define the normal curvature as
\begin{gather}
    F_{\alpha\beta}^{ij} = \partial_\alpha A_\beta^{ij} - \partial_\beta A_\alpha^{ij} + A_{\alpha}^{ik} A_{\beta,k}^j - A_{\alpha}^{jk} A_{\beta,k}^i. \label{eqapp:gausstorsion}
\end{gather}

$\Gamma, K$ and $A$ are not completely independent of each other. They are connected by the Gauss-Codazzi-Ricci equations:
\begin{gather}
    \nabla_{\alpha} K^i_{\beta\gamma} - \nabla_{\beta} K^i_{\alpha\gamma} = K^j_{\beta\gamma} A^i_{j,\alpha} - K^j_{\alpha\gamma} A^i_{j,\beta}, \notag\\
    \bar{R}_{\alpha\beta\gamma\delta} = K^i_{\beta\delta}K^i_{\alpha\gamma} - K^i_{\beta\gamma}K^i_{\alpha\delta} + R_{\alpha\beta\gamma\delta}, \notag\\
    F^{ij}_{\alpha\beta} = K^i_{\alpha\gamma}K^j_{\delta\beta}h^{\gamma\delta} - K^j_{\alpha\gamma}K^i_{\delta\beta}h^{\gamma\delta}, \label{eq:GCR}
\end{gather}
where we have used the Riemannian tensor of the induced metric $\bar{R}_{\alpha\beta\gamma\delta}$ and of the ambient space $R_{\alpha\beta\gamma\delta}$.

The Bonnet theorem establishes a one-to-one correspondence between the objects introduced above and immersions into $\mathbb{R}^n$. 
Here we will provide only the statement of the theorem:

{\it Suppose a metric tensor, a connection $A_{ij,\alpha}$ and a symmetric form $K^i_{\alpha\beta}$ are given in the simply connected domain $G$ of a Riemannian manifold $N^m$. Suppose these quantities satisfy $A_{ij,\alpha} = -A_{ji,\alpha}$, $K^i_{\alpha\beta}=K^i_{\beta\alpha}$, the Gauss-Codazzi-Ricci equations and are smooth. Then there exists a submanifold $F^m \subset \mathbb{R}^n$, unique up to a rigid motion, with induced metric coinciding with the given metric, having $K^i_{\alpha\beta}$ as its second fundamental form and $A_{ij,\alpha}$ as its normal connection. }



\section{Structure of the Bargman invariant (three-point function)}
\label{app:3pt}
A comprehensive discussion of properties of the Bargmann invariant is given in \cite{mukunda1993quantum}.

In comparison to the two-point function, which is real and only contains information about the distance between points, the three-point function provides information about the relative phase between states in a gauge invariant way (the relative phase between two points is gauge dependent, but for three states we can construct a gauge-invariant phase). 

We assume that $\arg \tr\left[P_1 P_2 P_3\right]  = -\int\limits_{\gamma} A$ for a certain path $\gamma$ in $\cpn$ that contains points $P_1, P_2$ and $P_3$ and try to determine which path reproduces gives the correct phase. We proceed as follows. For a fourth point $P$ we want to solve the equation
\begin{gather}
    \arg \tr\left[P_1 P P_2 P_3\right] = \arg \tr\left[P_1 P_2 P_3\right],
\end{gather}
which correspond to $P$ lying on the segment of the path between $P_1$ and $P_2$.

Using the reduction equation Eq. \eqref{eq:red} we get
\begin{gather}
      \arg \tr\left[P_1 P_2 P_3 \right] = \arg \tr\left[P_1 P P_2  P_3 \right] = \arg \tr\left[P_1 P_2 P_3\right] + \arg\tr\left[P_1 P P_2\right] \Rightarrow \arg\tr\left[P_1 P P_2\right] = 0
\end{gather}
where we have taken into account that $\tr[P_1 P_2]$ is a real number. Next, we represent $P_i$ and $\ket{u_i} \bra{u_i}$ for some states $\ket{u_i}$. We have
\begin{gather}
    \tr\left[P_1 P P_2\right] = \braket{u_1|u} \braket{u|u_2} \braket{u_2|u_1}, \quad r =\braket{u_2|u_1},
\end{gather}
and, without loss of generality, we choose the representatives $\ket{u_{1,2}}$ such that the $r$ is real. We can expand $|u\rangle = \alpha |u_1\rangle + \beta |u_2\rangle$, with $\left|\alpha\right|^2 + \left|\beta\right|^2 = 1$. Then,
\begin{gather}
     \tr\left[P_1 P P_2\right] = \braket{u_1|u} \braket{u|u_2} r = \left(\alpha + r \beta\right) \left(r \alpha^* + \beta^*\right) r = \left(r|\alpha|^2 + r \left|\beta\right|^2 + r^2\alpha^* \beta + \alpha \beta^*\right) r = \notag\\
     r^2 + r^3 \alpha^* \beta + \alpha\beta^* r \in \mathbb{R}.
\end{gather}
Finally, we arrive at
\begin{gather}
  \alpha^*\beta \in\mathbb{R}
\end{gather}
which implies $\alpha = \tan t \beta$, for some real $t$. Then the part of the path that connects $\ket{u_{1}}$ $\ket{u_{2}}$ is
\begin{gather}
    \ket{u(t)} = e^{i\phi}\left(\cos t \ket{u_1} + \sin t \ket{u_2}\right),
\end{gather}
and disregarding the common phase, we recognize $\ket{u(t)}$ as the geodesic in $\mathbb{C}P^n$  connecting the states $\ket{u_{1,2}}$. This allows one to conclude that
\begin{gather}
    \arg \tr[P_1 P_2P_3] = -\int\limits_{\gamma_{123}} A, \quad \gamma_{123}=\gamma_{12}+\gamma_{23}+\gamma_{31}.
\end{gather}

\section{Local expansion of the Bargman invariant} \label{app:geodesics}
\label{app:phase_geometry}

We begin by studying the geodesic distance between two points on a Riemannian manifold $(M,g_{ab})$ as a function of its metric. 
Let us consider some point $Z_0 \in M$ and pick coordinates $z^a$ such that the point $X_0$ correspond to the origin $z^a=0$. We expand the metric in the vicinity of $Z_0$:
\begin{gather}
    g_{ab}(z) = g_{ab} + z^c \partial_c g_{ab} + \frac12 z^c z^d \partial_c \partial_d g_{ab} + \cdots.
\end{gather}
The geodesic distance between the origin and $z$ $L(z)$ can be computed by solving the Hamilton-Jacobi equation for $\sigma(z)= L^2(z)$ \cite{dubrovin1992modern}:
\begin{gather}
    g^{ab} \frac{\partial \sigma(z)}{\partial z^a} \frac{\partial \sigma(z)}{\partial z^b} = 4\sigma(z).
\end{gather}
Solving this equation perturabitvely to the fourth order, we obtain
\begin{gather}
    L^2(z) = g_{ab} z^a z^b + \frac12\partial_a g_{bc} z^a z^b z^c + \frac16\left(\partial_a \partial_b g_{cd} -\frac12 \Gamma_{eab}g^{ef}\Gamma_{fcd}\right) z^a z^b z^c z^d + O(z^5). \label{eq:general_dist}
\end{gather}
We turn now to considering a submanifold $N$ immersed in a manifold $M$. For a point $X_0$ in $M$ that is also belongs to $N$ we introduce coordinates $(z_\alpha,z_i)$ with $z_\alpha$ parametrizing the submanifold and $z_i$ the normal dimensions. The submanifold $N$ is locally defined by $z_i=0$. In this case Eq. \eqref{eq:general_dist} can be used both for the "intrinsic" distance $L_N$ where geodesics are constrained $\gamma_N \subset N$ and the global distance with unconstrained geodesics $\gamma_M \subset M$. In the former case the induced versions of $g$ and $\Gamma$ should be used. One sees that the first three terms in Eq. \eqref{eq:general_dist} are the same in both cases, and only the last term can differ. In particular, the difference reads
\begin{gather}
    \Delta L^2 = L^2_{N} - L^2_{M} = \left(\frac{1}{12} g^{ij}\Gamma_{i,\alpha\beta}\Gamma_{j,\gamma\delta} + \frac{1}{6} g^{i\epsilon}\Gamma_{i,\alpha\beta}\Gamma_{\epsilon,\gamma\delta}\right) z^\alpha z^\beta z^\gamma z^\delta,
\end{gather}
the structure of this expression is elucidated when we introduce 
\begin{gather}
(\boldsymbol{K}_{\alpha\beta})^{i} = \Gamma_{\alpha\beta}^i, \quad (\boldsymbol{t}_{\alpha})^{i} = g_{\alpha i}, \quad (\boldsymbol{m})^{i j} = g^{i j},
\end{gather}
where $\boldsymbol{K}_{\alpha\beta}$ is the second fundamental from Appendix \ref{app:extrinsicgeometry} and bold symbols reflect the presence of normal indices. With the use of these object, we obtain
\bea \label{eq:delta_L}
\Delta L^2 = \frac{1}{12}  \left( \boldsymbol{K}_{\alpha \beta}^{T} \boldsymbol{m} \boldsymbol{K}_{\rho \tau} + 2 \boldsymbol{t}_{\sigma} \cdot \boldsymbol{K}_{\rho \tau} \Gamma^{\sigma}_{\alpha \beta}  \right) z^{\alpha} z^{\beta} z^{\rho} z^{\tau}.
\eea
It is a known fact \cite{Aminov2001} that when $\boldsymbol{K}_{\alpha \beta}$ vanishes geodesics stay within the submanifold and, in accordance with that, Eq. \eqref{eq:delta_L} implies that the global and intrinsic distances coincide.

After having covered the expansion of the geodesic distance, we wish to describe the phase that comes from integrating the two-form over geodesic triangles. To this end, we first need to derive a perturbative expression for the shape of the geodesic connecting two points and not just its length. The geodesic equation reads 
\bea
\frac{d^2 z^{a}}{ds^2} = \Gamma^{a}_{b c} \frac{dz^b}{ds} \frac{dz^c}{ds}.
\eea
We begin with solving for the geodesic with a specified initial condition $\left.\frac{dz^{a}}{ds}\right|_{s=0} = v^{a}$:

\bea
z^{a}(s) = v^{a} s - \Gamma_{\alpha \beta}^{a} v^{b} v^{c} \frac{s^2}{2} + \left(2 \Gamma_{b e}^{a} \Gamma^{e}_{c e} - \partial_{d}\Gamma^{a}_{b c} \right) v^{b} v^{c} v^{d} \frac{s^3}{6} + O(s^4),
\eea
where $s$ is the parameter along the geodesic.

Now we would like to relate the initial conditions vector $a^{\mu}$ to  and final point $z^{\mu}(1)=z^\mu$:
\bea
z^{a} = v^{a} - \frac{1}{2} \Gamma_{b c}^{a} v^{b} v^{c}  + \frac{1}{6}\left(2 \Gamma_{b e}^{a} \Gamma^{e}_{c d} -  \partial_{d}\Gamma^{a}_{b c} \right) v^{b} v^{c} v^{d} + O(a^4).
\eea 
Inverting this relation yields
\bea
v^{a} = z^{a} + \frac{1}{2}\Gamma^{a}_{bc} z^{b} z^{c} + \frac{1}{6} \left( \Gamma_{d e}^{a} \Gamma^{e}_{b c} + \partial_{b} \Gamma^{a}_{c d} \right) z^{b} z^{c} z^{d} + O(z^4).
\eea
Now we are finally ready to write down the shape of the geodesic in terms of just the final point $z^{\mu}$:
\begin{gather}
z^{a}(s) = \left(z^{a} + \frac{1}{2}\Gamma^{a}_{bc} z^{b} z^{c} + \frac{1}{6} \left( \Gamma_{b e}^{a} \Gamma^{e}_{c d} + \partial_{d} \Gamma^{a}_{b c} \right) z^{b} z^{c} z^{d}\right) s- \Gamma_{bc}^{a} \left( z^{b} z^{c} + z^{c} \Gamma_{d e}^{c} z^{d} z^{e} \right) \frac{s^2}{2} + \notag \\
+ \left(2 \Gamma_{b e}^{a} \Gamma^{e}_{c d} - \partial_{d}\Gamma^{a}_{bc} \right) z^{b} z^{c} z^{d} \frac{s^3}{6} + O(s^4). \label{eq:geod3rd}
\end{gather}
Next, consider the geodesic triangle formed by the origin, $z_1$ and $z_2$, the phase associated to it is given by
\bea
-\arg \text{tr}[P(0)P(z_1)P(z_2)] = \int\limits_{0}^{1} ds \left( \frac{d z_1^{a}}{d s}A_a(z_1(s)) + \frac{d z_{12}^{a}}{d s}A_a(z_{12}(s)) - \frac{d z_2^{a}}{d s}A_a(z_2(s)) \right),
\eea
where $z_i(s)$ is the geodesic that connects the origin and $z_i$ and $z_{ij}(s)$ is the geodesic that connects $z_i$ and $z_j$. To the leading order we obtain
\begin{gather}
-\arg \text{tr}[P(0)P(z_1)P(z_2)] =  \frac{1}{2} \omega_{a b} z_1^a z_2^b +
\frac{1}{6} \nabla_{a}\omega_{b c} (z_1^{a} z_1^{b} z_2^{c} + z_2^{a} z_1^{b}z_2^{c})  
+\frac{1}{4}\tilde{\Gamma}_{a b c} (z_1^{a} z_1^{b} z_2^{c} - z_2^{a} z_2^{b} z_1^{c}) + O(z^4),
\end{gather}
here $\nabla_{a}\omega_{b c} $ is the covariant derivative of $\omega_{a b}$ for which, in $\cpn$, we automatically have $\nabla \omega =0$ and  $\tilde{\Gamma}_{a, b c}(x) = \omega_{a d} \Gamma^{d}_{b c}$.

\section{Reconstruction of one-dimensional submanifolds} \label{app:reconstruct}
Here we will describe how a map $f: M_1 \to \mathbb{C}P^1$, where $M_1$ is a one-dimensional manifold, can be reconstructed from the knowledge of the induced metric $g$ and the 3-tensor $T$ on $M_1$, both of which in this case are represented by single numbers.

For reference, the metric, symplectic form and non-zero Christiffel tensors on $\mathbb{C}P^1$ are
\begin{gather}
    ds^2 = \frac{dzd\bar{z}}{\left(1+\left|z\right|^2\right)^2}, \quad \omega = -\sqrt{g} dz \wedge d\bar{z},\quad \Gamma^z_{zz} = - \frac{2\bar{z}}{1+\left|z\right|^2},\quad \Gamma^{\bar{z}}_{\bar{z}\bar{z}} = \left(\Gamma^z_{zz}\right)^*,\quad z = x+ i y.
\end{gather}
After choosing a real coordinate t on $M_1$ and a complex coordinate on $\mathbb{C}P^1$, $f(t)$ can be though as a complex-valued function of a real number. We later convenience we reparametrize $t$ so that
\begin{gather}\label{eq:fp_condition}
    \frac{\left|f'(t)\right|^2}{\left(1 + \left|f(t)\right|^2\right)^2} = 1,
\end{gather}
and, hence, $f'(t)$ gives a unit tangent vector at $f(t)$. This reparametrization is equivalent to $g_{tt}=1$ and, therefore, only require the knowledge of $g$. We now compute the convariant derivative of the tangent vector as
\begin{gather}
    n(t) = \nabla_{f'(t)} f'(t) =\left( f''(t) - \frac{2\bar{f}(t)}{1+\left|f(t)\right|^2} f'(t)^2, \bar{f}''(t) - \frac{2 f(t)}{1+\left|f(t)\right|^2} \bar{f}'(t)^2\right).
\end{gather}
One can check that $n(t)$ is perpendicular to $f(t)$ at each point of the line $M_1$. The vector $n(t)$ is in fact a symmetric tensor of rank 2 that represents the second fundamental form (see Appendix \ref{app:extrinsicgeometry} for definition). As the next step, we compute the tensor $T$ as follows
\begin{gather}
    T(t) = \omega(f'(t),n(t)) = \operatorname{Im}\left[\frac{2f''(t) \bar{f}'(t)}{\left(1+\left|f(t)\right|^2\right)^2} - \frac{4\bar{f}(t) f'(t)^2 \bar{f}'(t)}{\left(1 + \left|f(t)\right|^2\right)^3}\right] = \operatorname{Im}\left[ \frac{2 f''(t)}{f'(t)} - \frac{4\bar{f}(t) f'(t)}{(1+\left|f(t)\right|^2)} \right]. \label{eq:tcan}
\end{gather}
When $T(t)$ is known, this can be thought of as an equation on $f(t)$. Let us show how one can solve it. First, we represent $f$ as $f(t) = \tan \theta (t) e^{i \phi(t)}$. With that, the condition Eq. \eqref{eq:fp_condition} translates into
\bea
\phi'(t) = 2\frac{\sqrt{1-\theta'(t)^2}}{\sin 2\theta(t)}
\eea
Finally, Eq. \eqref{eq:tcan} becomes
\begin{gather}
T(t)=-\frac{2 \left(\theta ''(t)+2 \left(\theta '(t)^2-1\right) \cot (2 \theta(t))\right)}{\sqrt{1-\theta '(t)^2}},
\end{gather}
which is a simple differential quation the solution of which uniquely restores the map $f(t)$ from $M_1$ to $\mathbb{C}P^1$. 

\section{Two surfaces in $\mathbb{C}P^{3}$ with the same $\omega$ and $g$ but different extrinsic geometries \label{App:example}}
In this Appendix we will construct a family of two-dimensional submanifolds of $\mathbb{C}P^3$ that have the same induced metric and Berry curvature, but different  extrinsic geometry and the three-tensors. We start by reviewing the Veronese immersion $\mathbb{C}P^1 \times \mathbb{C}P^1 \to \mathbb{C}P^3$ \cite{harris2013algebraic}. Let $z_{1,2,3}$ be coordinates on $\mathbb{C}P^3$ and $s,t$ be holomorphic coordinates on the two copies of $\mathbb{C}P^1$. We consider the following immersion defined by two complex coordiantes $(s,t) \in \mathbb{C}^2$:
\begin{gather}
   z_1 = s,\quad z_2=t,\quad z_3=st.
\end{gather}
One can see that, in this case, the induced metric and curvature form are
\begin{gather}
    ds^2 = \frac{|ds|^2}{(1+|s|^2)^2}+ \frac{|dt|^2}{(1+|t|^2)^2},\quad \omega= \frac{ds\wedge d\bar{s}}{(1+|s|^2)^2} + \frac{dt\wedge d\bar{t}}{(1+|t|^2)^2},
\end{gather}
which are just the metric and curvature from of two copies of $\mathbb{C}P^1$. 

Next, let us consider two real coordinates $(x,y)$ on a 2-torus $\mathbb{T}^2 = S^1\times S^1$ and maps $s=f(x) , f(0)=f(1)$ and $t=g(y), g(0)=g(1)$, such that

\bea \label{eq:fp_cond}
\frac{|f'|^2}{(1+|f|^2)^2} = \frac{|g'|^2}{(1+|g|^2)^2}=1.
\eea
This defines a map $\mathbb{T}^2 \to \mathbb{C}P^1 \times \mathbb{C}P^1 \to \mathbb{C}P^3$. This map leads to the following induced metric and 2-form:
\begin{gather}
    ds^2 = dx^2 + dy^2, \quad \omega =0.
\end{gather}
We note that since $\omega=0$ all Berry phases for contractable loops are equal to zero. There are two independent non-contractable loops --- the one that goes along $x\in \left[0,1\right]$, $y=0$ and the other $y\in\left[0,1\right]$, $x=0$. In this case the Berry phases are known as the Zak phases and are equal to
\begin{gather}
    \phi_f =\frac12 \int\limits^1_0 \frac{f \bar{f}' - \bar{f} f' }{1+\left|f\right|^2} dx, \quad  \phi_g =\frac12 \int\limits^1_0 \frac{g \bar{g}' - \bar{g} g' }{1+\left|g\right|^2} dx.
\end{gather}
Meanwhile, the three tensor is
\begin{gather}
    T:\quad T_{xxx}(x) = \frac{2 f''(x)}{f'(x)} - \frac{4 \bar{f}(x) f'(x)} {(1+|f(x)|^2)}, \quad T_{yyy}(y) = \frac{2 g''(y)}{g'(y)} - \frac{4 \bar{g}(y) g'(y)} {(1+|g(y)|^2)}.
\end{gather}
One sees that by changing $f,g$ one can change $T$, but the metric and curvature form are the same.

It is also convenient to have an expression for the 3-tensor, when the condition Eq. \eqref{eq:fp_cond} is not enforced:
\begin{gather}
    T_{xxx}(x) =  \left|f'(x)\right|^2 \operatorname{Im}\left[ \frac{2 f''(x)}{f'(x)} - \frac{4\bar{f}(x) f'(x)}{(1+\left|f(x)\right|^2)} \right]
\end{gather}
and similarly for $T_{yyy}$.

A Hamiltonian that has this state as an eigenstate can be constructed quite simply: for any two complex functions of real arguments $f(x), g(y)$ consider
\begin{gather} \label{eq:veronese_ham}
    H = O_{x,y} \left(\mathbf{1}_{4} - \psi_{f,g} \psi_{f,g}^\dagger\right), \quad \psi_{f,g} = \frac{1}{\sqrt{1+|f(x)|^2} \sqrt{1+|g(y)|^2}} \begin{pmatrix}
    1\\
    f(x)\\
    g(y)\\
    f(x) g(y)
    \end{pmatrix},
\end{gather}
where $O_{x,y}$ is an arbitrary hermitian operator, that commutes with the projector $\left(\mathbf{1}_{4} - \psi_{f,g} \psi_{f,g}^\dagger\right)$. 

Hamiltonian Eq. \eqref{eq:veronese_ham} can be coming from a local hopping Hamiltonian on a lattice if we choose
\begin{gather}
f(k_x) = \tan k_x e^{i  n  k_x},\quad g(k_y) = \tan k_y e^{i  m  k_y}.
\end{gather}
Explicitly, the wave function is
\begin{gather}
    \psi(k) = \begin{pmatrix}
    \cos k_x \cos k_y \\
    \sin k_x \cos k_y e^{i n k_x}\\
    \cos k_x \sin k_y e^{i m k_y}\\
    \sin k_x \sin k_y e^{i n k_x + i m k_y}
    \end{pmatrix},
\end{gather}
Then the Hamiltonian is
\bea
H(k) = \psi(k) \psi^\dagger(k) = \begin{pmatrix}
 c_x^2 c_y^2 & c_x c_y^2 s_x e^{-i k_x n} &
   c_x^2 c_y s_y e^{-i k_y m} & c_x c_y s_x
   s_y e^{-i k_x n-i k_y m} \\
 c_x c_y^2 s_x e^{i k_x n} & c_y^2 s_x^2 & c_x
   c_y s_x s_y e^{i k_x n-i k_y m} & c_y s_x^2
   s_y e^{-i k_y m} \\
 c_x^2 c_y s_y e^{i k_y m} & c_x c_y s_x
   s_y e^{i k_y m-i k_x n} & c_x^2 s_y^2 & c_x
   s_x s_y^2 e^{-i k_x n} \\
 c_x c_y s_x s_y e^{i k_x n+i k_y m} & c_y
   s_x^2 s_y e^{i k_y m} & c_x s_x s_y^2 e^{i
   k_x n} & s_x^2 s_y^2 \\
\end{pmatrix}
\eea
where $c_{x/y} = \cos k_{x/y}$ and $s_{x/y} = \sin k_{x/y}$.

\section{Response to an Inhomogenous Electric Field} \label{app:conduct}
Closely following \cite{Kozii_2021}, we will study the current response to an external vector potential $A^{\alpha}(q,\omega) e^{i \omega t + i q x}$:
\bea
j_{\alpha}(q,\omega) = T_{\alpha\beta}(q,\omega) A^{\beta}(q,\omega),
\eea
where $j_{\alpha}(q,\omega)$ is the current Fourier component and $T_{\alpha\beta}(q,\omega)$ is the tensor to be determined. We can introduce an electric field by $E^{\alpha} = i \omega A^{\alpha}$, leading to
\bea
j_{\alpha}(q,\omega) = \sigma_{\alpha\beta}(q,\omega) E^{\beta}(q,\omega), ~ \sigma_{\alpha\beta}(q,\omega) = T_{\alpha\beta}(q,\omega)/(i \omega),
\eea
where $\sigma_{\alpha\beta}(q,\omega)$ is the conductivity tensor.

The Kubo formula yields
\bea \label{eq:T_def}
 T_{\alpha\beta}(q,\omega) = -\sum_{n,m} \int \frac{d^d k}{(2\pi)^d} \frac{n(\epsilon_n(k)) - n(\epsilon_m(k+q))}{\epsilon_n(k) - \epsilon_m(k+q) + i \omega} F^{nm}_{\alpha \beta} (k ,q),
\eea
where $n(\epsilon)$ is the electron distribution function and
\bea
F^{nm}_{\alpha \beta} (k ,q) = \tr[P^{(n)}(k) j^{\alpha}(k+q/2)P^{(m)}(k+q)j^{\beta}(k+q/2)]
\eea
with $P^{(n)}(k)$ being the projector on the n'th band and $j^{\alpha}(k) = \frac{\partial H(k)}{\partial k^{\alpha}}$ being the current operator.

In order to obtain an answer where the response is determined solely by geometry we consider an insulator at zero temperature. We additionally require that only one lowest band is filled. Furthermore, we assume that this band is separated from all other bands by the gap $\Delta$, all bands are flat and all unoccupied bands are at the same energy.

In this case, the integrand in Eq. \eqref{eq:T_def} becomes
\begin{gather}
\frac{\tr[P(k) j^{\alpha}(k+q/2)j^{\beta}(k+q/2)]}{-\Delta + i \omega} + \frac{\tr[P(k) j^{\beta}(k-q/2)j^{\alpha}(k-q/2)]}{-\Delta - i \omega}-\notag\\
-  \frac{\tr[P(k) j^{\alpha}(k+q/2) P(k+q) j^{\beta}(k+q/2)]}{-\Delta + i \omega} - \frac{\tr[P(k) j^{\beta}(k-q/2)P(k-q)j^{\alpha}(k-q/2)]}{-\Delta - i \omega},
\end{gather}
where $P(k)$ is the projection on the occupied band. In the limit of a constant electric field $\omega  = 0$, we obtain for the conductivity
\begin{gather} \label{eq:sigma_expr}
\sigma_{\alpha\beta}(q) = i \sum_{n,m} \int \frac{d^d k}{(2\pi)^d} \left( \tr[P(k) \partial_\beta P(k-q/2) \partial_\alpha P (k-q/2)] - \tr[P(k) \partial_\alpha P(k+q/2)\partial_\beta P(k+q/2)] \right).
\end{gather}
This expression is composed of products of projectors and theirs derivatives. Let us show how all of them can be written with the use of $\mathcal{A}^x_\alpha$ defined in Eq. \eqref{eq:uglyA}. We find it useful to use the  following expressions for $\mathcal{A}^x_\alpha$ and $\bar{\mathcal{A}}^x_\alpha$
\begin{gather}
    \mathcal{A}^x_\alpha(y) = \frac{\tr[P(x)P(y)\partial_\alpha P(y)] }{\tr[P(x)P(y)]}, \quad  \bar{\mathcal{A}}^x_\alpha(y)  = \frac{\tr[P(x) \partial_\alpha P(y) P(y)] }{\tr[P(x)P(y)]}.  
\end{gather}
First of all, using the identity $P^2=P$ and decomposition formula Eq. \eqref{eq:red} we can compute the derivative
\begin{gather}
    \partial_{z_\alpha} \tr[P(x)P(y)P(z)]  = \tr[P(x)P(y) P(z)\partial_\alpha P(z)]+ \tr[P(x)P(y)\partial_\alpha P(z) P(z)] = \notag\\
    = \frac{\tr[P(x)P(y) P(z)] \tr[P(x)P(z)\partial_\alpha P(z)]}{\tr[P(x)P(z)]} + \frac{\tr[P(x)P(y) P(z)] \tr[P(y)\partial_\alpha P(z)P(z)]}{\tr[P(y)P(z)]} = \notag\\
    = \left(\mathcal{A}_\alpha^x(z) +\bar{\mathcal{A}}^y_\alpha(z)\right) \tr[P(x)P(y)P(z)]  \label{eq:gender3pt}.
\end{gather}
Next, using similar manipulations for the tensor that appears in Eq. \eqref{eq:sigma_expr}, we obtain
\begin{gather}
    \tr[P(x)\partial_\alpha P(y) \partial_\beta P(y)] = \tr[P(x) P(y) \partial_\alpha P(y) \partial_\beta P(y)] + \tr[P(x) \partial_\alpha P(y) P(y) \partial_\beta P(y)] = \notag\\
    =  \tr[P(x) P(y) \partial_\alpha P(y) \partial_\beta P(y) P(y)] + \tr[P(x) \partial_\alpha P(y) P(y) \partial_\beta P(y)] = \notag\\
   \tr[P(x)P(y)] \tr[P(y) \partial_\alpha P(y) \partial_\beta P(y)] + \frac{\tr[P(x) \partial_\alpha P(y) P(y)] \tr[P(y) \partial_\beta P(y) P(x)]}{\tr[P(x) P(y)]} = \notag\\
  =\tr[P(x)P(y)] \left(\bar{\mathcal{A}}_\alpha^x(y) \mathcal{A}_\beta^x(y) + Q_{\alpha\beta}(y) \right),
\end{gather}
where we have used $P^2=P$, $P\partial_\alpha P P = 0$ and the reduction formula Eq. \eqref{eq:red}.

Finally, we arrive at   
\begin{gather*}
\sigma_{\alpha\beta}(q) = i\int \frac{d^dk}{(2\pi)^d} \left(P(k,k+q/2) \left(\mathcal{A}^{k + \frac{q}{2}}_{\alpha}(k) \bar{\mathcal{A}}^{k+ \frac{q}{2}}_{\beta}(k) + \bar{Q}_{\alpha \beta}(k)\right)-P(k,k-q/2) \left(\bar{\mathcal{A}}^{k -  \frac{q}{2}}_{\alpha}\left(k \right) \mathcal{A}^{k - \frac{q}{2}}_{\beta}\left(k\right) + Q_{\alpha \beta}(k) \right) \right).  
\end{gather*}
Expanding to the second order in $q$ yields
\begin{gather}
\sigma_{\alpha\beta}(q) = -\int \frac{d^d k}{(2\pi)^d} \left(\omega_{\alpha \beta} - (\omega_{\alpha \beta} g_{\gamma \delta} + g_{\alpha \gamma} \omega_{\beta \delta} + \omega_{\gamma \alpha} g_{\beta \delta} ) \frac{q^{\gamma} q^{\delta}}{4} \right).
\end{gather}

\section{Modern Theory of Polarization} \label{app:polarization}
One of the application of the developed theory of extrinsic geometries of the bands and the particular application of the functions $P$ could serve the theory of polarization.Thus, the third gauge-invariant cumulant of electron configuration was introduced in \cite{SouzaWilkensMartin2000} and could be expressed with the use of the extrinsic three-tensor. The generating function for gauge-invariant cumulatns of higher orders could be written in the following form,

\bea
\log C(q) = \frac{V}{(2 \pi)^3} \int dk \log \langle u_k | u_{k + q} \rangle,
\eea
where V is the volume of the system.
Let us note that the right hand side of the above equation is defined up to a factor of $2\pi i n$, making $C(\alpha)$ a nice gauge invariant function.
The cumulants are defined by differentiating the logarithm of the generating function with respect to $\alpha$
\bea \notag
\langle\hat{X}^i \dots \hat{X}^j \rangle = i \partial_{q_i} \cdots \partial_{q_j} \log C(q)|_{q = 0}.
\eea

For the first few cumulants we obtain:

\bea
A^i = \langle \hat{X}^i \rangle = \frac{V}{(2 \pi)^2}  \int_{BZ} dk ~ A^i(k),~ \langle \hat{X}^i \hat{X}^j \rangle = \frac{V}{(2 \pi)^2}  \int_{BZ} dk ~ g^{i j}(k),~ \langle \hat{X}^i \hat{X}^j \hat{X}^k \rangle = -\frac{V}{(2 \pi)^2} \int_{BZ} dk ~ T^{i j k} (k).
\eea
We will show how this generating function could be computing with the use of the n-point functions $P$. We note that for small $\alpha$ we have
\begin{gather}
    \log C(q) \approx \vec{q} \vec{A} + \ldots \label{eq:smallexpan}
\end{gather}
For general $\alpha$ and $\beta$ we can show that 
\bea
\log \frac{C(\alpha + \beta)}{C(\alpha) C(\beta)} = \int dk \left[ \log P^{(2)}(k, k+\alpha+\beta) - \log P^{(3)}(k, k+\alpha, k+\alpha+\beta) \right]. \label{eq:Avdrel}
\eea
Now let us define $b_{\vec{v}}(t) = \log C(\vec{v} t)$. Using the equations \eqref{eq:Avdrel} and \eqref{eq:smallexpan} for $\alpha = \vec{v} t$ and $\beta= \vec{v} \epsilon$ and expanding up to first order in $\epsilon$ we get
\bea
b_{\vec{v}}'(t) - \vec{v} \cdot \vec{A} =  \frac{V}{(2\pi)^2} v^{\alpha} \int dk \left[ \mathcal{A}^{k}_{\alpha}(k + t v)  - 2 u_{\alpha}^{k}(k + v t)  \right],
\eea
Integrating this equation we arrive at
\bea
\log C(q) = \vec{\alpha} \cdot \vec{A}+  \frac{V}{(2\pi)^2} q^{\alpha} \int d^2 k \int_{0}^{1} dt \left[ \bar{\mathcal{A}}^{k}_{\alpha}(k + t \vec{q})  \right].
\eea

\section{Intersection Number} \label{app:intersection}
As has been shown elsewhere in the text, in order to fully characterize the immersion of a parameter space $M$ into the complex projective space $\cpn$, along with intrinsic properties of the immersion, the Berry phase and quantum metric, one must also take into account extrinsic ones, for example, the three-tensor $T$. It is interesting to note, that in addition to local and continuously varying objects such as $T$, there are also global and discrete quantities that characterizes the immersion, namely, the self-intersection number or the Whittney invariant \cite{Aminov2001}. This quantity is stable under small smooth local deformations of the parameter space, but can be changed with transformation that violate smoothness.

This is best illustrated when we consider an immersion of the circle $S^1$ into $\mathbb{C}P^1$. In this case the self-intersection number can be computed as
\begin{gather}
    N_{SI} = \int dt ~ \epsilon_{\mu\nu} v^\mu v^\rho \nabla_{\rho} v^\mu, \quad v^\mu = \frac{dz^\mu}{dt},
\end{gather}
where $z(t), z(0)=z(1), t \in [0,1]$ parametrizes the immersion ($z$ is a complex coordinate on $\mathbb{C}P^1$), and $\nabla$ is the covariant derivative.

The simplest example of a self-intersecting map is the "figure $8$" immersion defined by $z(t) = \sin t e^{i t}$. If we only allow smooth changes of this curve the self-intersection cannot be removed. The self-intersection can by removed by a continuous, but non-smooth transformation, although the derivative $z'(t)$ will become singular at some point in the process \cite{arnold1994topological}.

Similarly, the self-intersection invariant $N_{SI}$ can be defined for maps $M_2 \to \mathbb{C}P^2$ (and, generally, $M_n \to \mathbb{C}P^n$). The expression is particularly simple, when written in terms of the normal curvature or Gauss torsion \eqref{eqapp:gausstorsion}. In two dimensions, we always have  $F^{ij}_{\alpha\beta} = \epsilon^{ij} F_{\alpha\beta}$, which allows us to write
\begin{gather}
N_{SI} = \int d^2 x ~ \epsilon^{\alpha \beta} F_{\alpha\beta}.
\end{gather}

\section{Geometry of Landau Levels} \label{sec:GLL}
We start by briefly reviewing the wave functions of Landau Levels \cite{ozawa2021relations}. The Hamiltonian of a free particle in an external magnetic field is
\begin{gather}
   \hat{H} = \frac{\hat{p}_x^2}{2m} + \frac{\left(\hat{p}_y - B x\right)^2}{2m},
\end{gather}
where $\hat{p}_i$ is the momentum operator. Using $p_y$ and the oscillator level $m$ as quantum numbers, we write the spectrum as
\begin{gather}
    \chi_{m,k_y}(x) = h_m\left(x - \frac{k_y}{B}\right) = \exp\left[ - \frac{B}{2}\left(x-\frac{k_y}{B}\right)^2 \right]  H_m\left[\sqrt{B} \left(x - \frac{k_y}{B}\right)\right], \label{eq:holfunLL}
\end{gather}
where $H_m$ is the Hermite polynomials and all functions at level $m$ have the same energy. The band wave functions are introduced as
\begin{gather}
    \psi_{m,k_x, k_y}(x ,y) = e^{- i \frac{k_x k_y}{2 B}}\sum \limits^\infty_{l=-\infty} e^{ i k_x a_x l + i \left(k_y + 2\pi  l/a_y\right) y}   H_m\left(x - a_x l - k_y/B\right), \label{eq:holfunLL}.
\end{gather}
The Bloch wave functions are
\begin{gather}
    u_{m,k}(x,y) = e^{- i \frac{k_x k_y}{2 B}} \sum \limits^\infty_{l=-\infty} e^{ -i k_x (x - a_x l) + 2\pi i l y/a_y}   H_m\left(x - a_x l - k_y/B\right), \quad u_k(x,y) = u_{0,k}(x,y), \label{eq:LL}
\end{gather}
where $a_x$ and $a_y$ are arbitrary numbers satisfying $a_x a_y B = 2\pi$. Restricting ourselves to the lowest Landau Level $m=0$, we now compute the overlap
\begin{gather}
    \braket{u_{k}|u_q} = \int\limits^{a_x}_0 dx \int \limits^{a_y}_0 dy u^*_{k}(x,y) u_q(x,y) 
    = a_y \int\limits^{a_x}_0 dx e^{i(k_x - q_x)(x-a_x l)} H_0(x - a_x l - k_y/B) H_0(x - a_x l -q_y/B) = \notag\\
    = a_y \int\limits^\infty_{-\infty} dx e^{i(k_x - q_x) x} H_0(x - k_y/B)H_0(x - q_y/B) =  a_y \int\limits^\infty_{-\infty} dx e^{i(k_x - q_x) x} \exp\left[ - \frac{B}{2}\left(x-\frac{p_y}{B}\right)^2 - \frac{B}{2}\left(x-\frac{k_y}{B}\right)^2\right] = \notag\\
    =\frac{\sqrt{\pi } \exp \left(-\frac{(\vec{k}-\vec{q})^2}{4 B}+\frac{ - i k_x q_y+ i q_x
   k_y}{2 B}\right)}{\sqrt{B}},
\end{gather}
which allows one to obtain the three-point phase
\begin{gather}
    \Phi(k,q) = \frac{\epsilon_{\alpha\beta}k^\alpha q^\beta}{2B},\quad \Phi(k,q,p) = \Phi(k,q)+\Phi(q,p)+\Phi(p,k) =  \frac{\operatorname{Area}(k,q,p)}{2B}, \label{eq:3ptLL}
\end{gather}
where $\operatorname{Area}$ is an area of a plane triangle that passes through the points $(k,q,p)$.

It can be also seen that the quantum metric is flat and the intrinsic quantum distance is $d_{\text{int}}(k,p)=\sqrt{\left(\vec{k}-\vec{p}\right)^2}$. Meanwhile, the global distance, while still a function of $\left(\vec{k}-\vec{q}\right)^2$, is different from the intrinsic one:
\begin{gather}
    \cos d_{\text{glob}}(k,q) =  \exp\left[ - \frac{\left(\vec{k} - \vec{q}\right)^2}{2B}\right].
\end{gather}
In fact, the phase of the three point function is the same for higher excited Landau levels \eqref{eq:3ptLL}. The quantum distance is different, but still depend only on the intrinsic distance.

\section{The Geometry of Twisted Bilayer Graphene in the Chiral Limit} \label{sec:TBG}
In this Appendix, we show that the wavefunctions of the lowest Landau level and the flat band of twisted Bilayer Graphene (tBLG) are holomorphic functions of the quasimomentum. Along with the Calabi rigidity this implies that any other band structure with the same metric $g$ and curvature $\omega$ (or even just $\omega$) will have the same extrinsic geometry as well. 

First, starting with the lowest Landau Level, we recast the expression Eq. \eqref{eq:LL}  as
\begin{gather}
    \psi_{0,k}(z) = \frac{1}{\vartheta\left(\frac{z}{a_x} + \frac{i k}{2 eB}; \tau \right)} \exp\left[\frac12 eB\left(\frac{z}{a_x} + \frac{i k a_x}{2 e B}\right)^2 + \frac{i}{2} \bar{k} z - \frac{1}{2 a_x^2} e B z\bar{z} \right], \notag
\end{gather}
where $z = x+ i y$ and $\tau = i \frac{a_y}{a_x}$. We see that the Bloch function
\begin{gather}
    u_{k}(z) = \psi_{0, k}(z) e^{- i\left(k,r\right)} = \frac{1}{\vartheta\left(\frac{z}{a_x} + \frac{i k a_x}{2 eB}; \tau \right)} \exp\left[\frac12 eB\left(\frac{z}{a_x} + \frac{i k a_x}{2 e B}\right)^2 - \frac{i}{2} k \bar{z} - \frac{1}{2 a_x^2} e B z\bar{z} \right] \label{eq:holfun}
    \end{gather}
    with the boundary conditions
    \begin{gather}
    u_{k}(z+a_x) = u_{k}(z) \exp\left[\frac{1}{2 a_x} e B z - \frac{1}{2 a_x} e B \bar{z}\right], \quad u_{k}(z+i a_y) =  u_{k}(z) \exp\left[\frac{1}{2 a_x} e B z \bar{\tau} - \frac{1}{2 a_x} e B \bar{z}\tau\right]
\end{gather}
is a holomorphic function of the quasi-momentum $k$: $\bar{\partial}_k u_{k}(z) = 0$. 

We proceed similarly, for the chiral limit of twisted bilayer graphene. In this case, as it was shown in \cite{tarnopolsky2019origin,popov2021hidden}, the Bloch wave function is
\begin{gather}
    v^\alpha_k(z,\bar{z}) \propto u_{k}(z) v^\alpha_0(z,\bar{z}),
\end{gather}
where $\alpha$ is the layer index, $u_k(z)$ is the Landau level Bloch function and $v^\alpha_0(z,\bar{z})$ is a more complicated function that was studies numerically in \cite{tarnopolsky2019origin}. Again we have that $\bar{\partial}_k v^\alpha_k(r) = 0$ and therefore the immersion into $\mathbb{C}P^{\infty}$ is holomorphic and all geometric properties can be recovered from the Berry curvature and quantum metric. 

\end{widetext}

\end{document}